\documentclass[12pt,letterpaper]{article}
\usepackage{amsmath}
\usepackage{amsfonts}
\usepackage[figurewithin=section,tablewithin=section]{caption}
\usepackage[usenames,dvipsnames]{color}
\usepackage{graphicx}
\usepackage{longtable}
\usepackage{rotating}
\usepackage[pdftex,bookmarks=false]{hyperref}
\hypersetup{pdfauthor={John Sibert}
            pdfsubject={Yellowfin stock assessment}
            pdftitle={Assessment of the yellowfin tuna Thunnus albacares 
stock in the Main Hawaiian Islands Yellowfin Tuna Fishery}
            pdfkeywords={yellowfin tuna,state space model,stock assessment,Hawaii}
            }%

\newcommand\singlespacing{\baselineskip=1.0\normalbaselineskip}
\renewcommand\deg[1]{$^\circ$#1}
\newcommand\SD{SEAPODYM}
\newcommand\MFCL{MULTIFAN-CL}

\newcommand\peryr{yr$^{-1}$}

\newcommand\MSY{\widetilde{Y}}
\newcommand\Fmsy{F_{\MSY}}

\title{Assessing of a portion of the Pacific 
{\it Thunnus albacares} stock:\\[0.125in]
{\it Ahi} in the Main Hawaiian Islands}

\author{
John Sibert\thanks{sibert@hawaii.edu}\\
Joint Institute of Marine and Atmospheric Research\\
University of Hawai`i at M\={a}noa\\
Honolulu, HI  96822 U.S.A.\\[0.125in]
\date{\today}
}

\begin{document}
\maketitle


\begin{abstract}
Regional tuna fishery management organizations cannot provide specific
advice to local fishery managers in small island jurisdictions. The
State of Hawaii maintains time series of yellowfin tuna catches dating
back to 1949, but these data have never been formally applied to
evaluating the effects of the yellowfin fishery in the Main Hawaiian
Islands on the local stock. I develop a new approach utilizing these
data that links the local stock dynamics to the dynamics of the larger
Pacific stock. This approach uses a state-space logistic production
model linked to the larger Pacific stock using an index of abundance. The
conclusion is that such a model is feasible, that the local stock
is not overfished and that local fisheries are fishing at acceptable
levels.
\end{abstract}

\section*{Introduction}
The responsibility to manage fisheries for tunas and tuna-like
species usually
lies with regional organizations established under international treaties.
In the Western Central Pacific Ocean (WCPO), this responsibility devolves to
the Western and Central Pacific Fisheries Commission (WCPFC, 
\url{https://www.wcpfc.int}).
The WCPFC conducts stock assessments for several species of tunas and
implements fishery management and conservation measures based on
these stock assessments. 
These WCPFC stock assessments and conservation and management measures
may assist the WCPFC in regulating large scale fisheries in the stock
habitat, but offer little useful advice to managers of small scale
fisheries operating outside of the primary equatorial fishing grounds.

\begin{figure}
\begin{center}
\includegraphics[width=\textwidth]{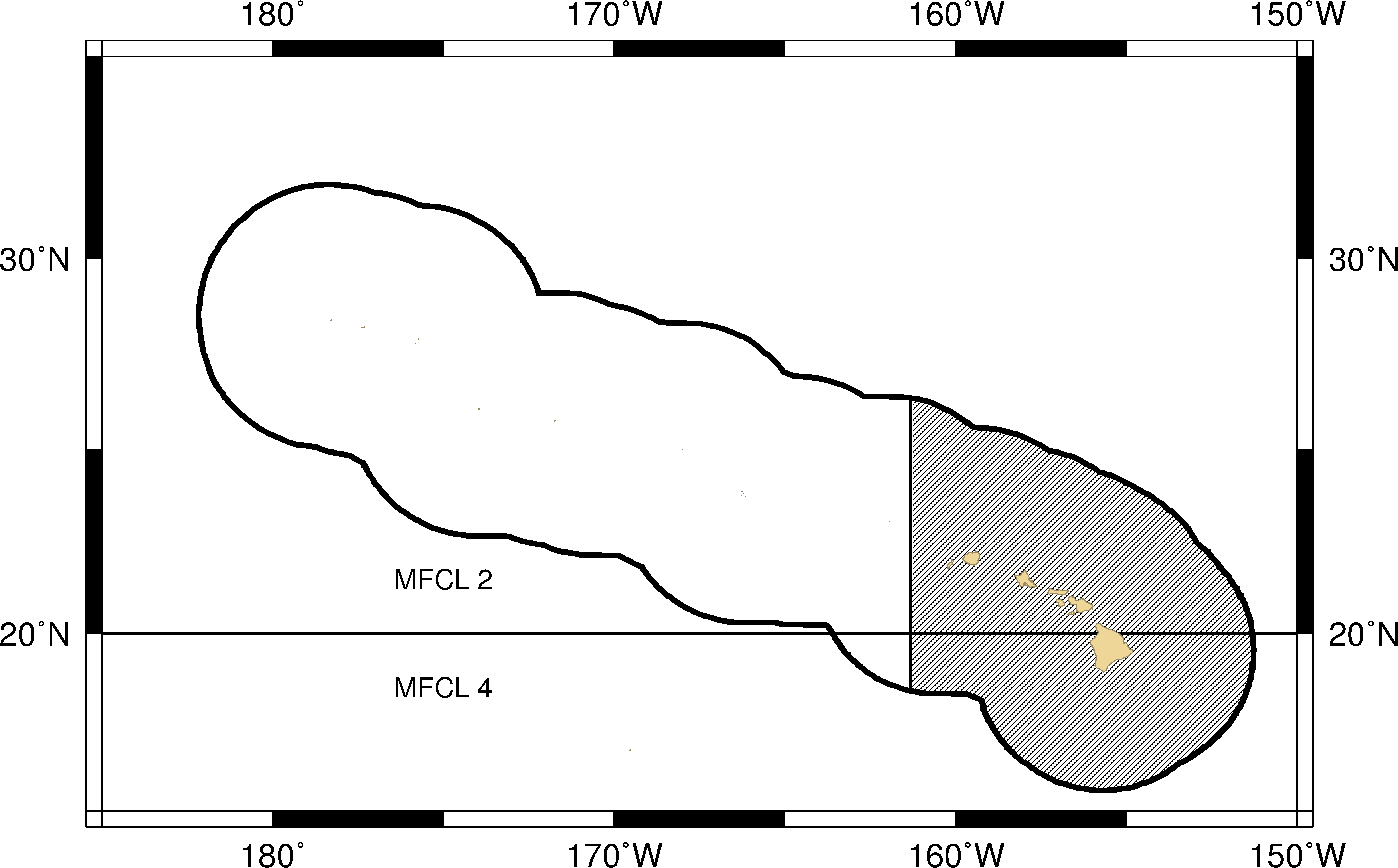}
\caption{\label{fig:mhimap}
The United States Exclusive Economic Zone around the Hawaiian
Archipelago. The Main Hawaiian Islands lie in the shaded area at the
extreme east of the EEZ. The line at 20\deg{N} latitude is the
boundary between WCPFC stock assessment regions 2 and 4. See text for
full explanation.
}
\end{center}
\end{figure}

Yellowfin tuna ({\it Thunnus albacares}, YFT) is an important food resource
of many Pacific Island communities and the people of the Hawaiian
islands of been fishing for yellowfin for many generations. 
Recent studies (Rooker et al. 2016) have shown that the Hawaii-based
fishery for YFT is  supported by
local production with little or no subsidies from equatorial
production zones or nurseries.
The Main Hawaiian Islands (MHI), Figure~\ref{fig:mhimap},
comprising the 8 largest islands in the Hawaiian Archipelago,
have been important fishing grounds throughout this history.
Since the recent expansion of the Papah\={a}naumoku\={a}kea Marine National
Monument in 2016, the MHI is the only legally accessible tuna fishing ground
in the State of Hawaii.
This change emphasizes the importance of careful management of yellowfin
fisheries in the MHI.

The primary purpose of this paper is to explore a means to apply the 65
year time series of
data collected from local fisheries to the problem of
providing advice to fisheries
managers about the effects of local fisheries on the local portion of
the larger stock of yellowfin tuna in the Pacific.
The general approach is to develop a relatively simple population
dynamics model, estimate the model parameters from existing
data, and apply the estimated parameters to answer the question of the
effects of the fishery.
I adopt a state-space variant of
the logistic surplus production model with several novel features.
Fishing mortality, the proportion of the population harvested each
year, is represented as a random walk of multiple gear types.
Offline coupling is used to link the local biomass to regional
biomass estimates by large scale yellowfin stock assessment models
developed for the WCPFC.
The logistic parameters of the surplus production model are
reparameterized to estimate parameters of direct relevance to fishery
management. All state transitions are represented as random effects.

\section*{Methods}
\label{sec:models}
State-space models separate variability in the biological
processes in the system (transition model)
from errors in observing features of interest
in the system (observation model).

The general form of the {\bf transition model} is
\begin{equation}
\alpha_t=T(\alpha_{t-1}) + \eta_t
\end{equation}
where $\alpha_t$ is the state at time $t$ and 
the function $T$ embodies the dynamics mediating the
development of the state at time $t$ from the state at the previous
time with random process error, $\eta_t$.

Stock dynamics follow the classic Schaefer (1954) differential equation:
\begin{equation}
\label{eqn:ischaefer}
\frac{dN}{dt} = rN(1-\frac{N}{K}) - FN
\end{equation}
where $N$ is the biomass of YFT in the MHI (measured in metric tonnes, mt), 
$r$ is the logistic growth rate (\peryr),
$K$ is the asymptotic biomass (mt), and
$F$ is the total fishing mortality (\peryr) in the MHI.

The state space transition equation is developed by solving
(\ref{eqn:ischaefer}) analytically from one time to the next and
applying a random error.
\begin{equation}
\label{eqn:intschaeferA}
N_t = \frac{K(r-\bar{F}_t)}{\frac{K(r-\bar{F}_t)}{N_{t-\Delta t}}e^{-\Delta
t(r-\bar{F}_t)}-re^{-\Delta t(r-\bar{F}_t)} -r} \cdot e^{\eta_t};
\quad \eta_t\sim N(0,\sigma^2_N)
\end{equation}
where $\eta_t \sim N(0,\sigma_N)$ is a process error expressing
variability in population dynamics
and $\bar{F}_t$ is the total fishing mortality exerted by all gears, i. e.,
\begin{equation}
\bar{F}_t =\sum_{g=1}^n F_{g,t}.
\end{equation}
The logarithm of fishing mortality for each gear is assumed to
follow a random walk with normal increments, as suggested by Nielsen
and Berg (2014).
\begin{equation}
\label{eqn:Fwalk}
\log F_{g,t} = \log F_{g,t-1} + \xi_t;\quad \xi_t\sim
N(0,\sigma^2_F)
\end{equation}
where  $\xi_t$ is a process error expressing the year to year
variability in fishing mortality.

An index of abundance assumes that the biomass of YFT in the MHI
is approximately proportional to some index which is independent of
the data analyzed in the model,
\begin{equation}
\log N_t - \log (Q\cdot I_t) = \omega_t;\quad \omega_t\sim N(0,\sigma^2_I)
\label{eqn:index}
\end{equation}
where
$I_t$ is the index (mt) at time $t$,
$Q$ is the estimated ratio of the MHI population size to the index,
and $\omega_t$ is a process error representing the difference
between the MHI biomass and the abundance index. 
Conventional fishery independent methods, such as acoustic or trawl
surveys, are not available for tunas. 
Biomass estimates from the most recent WCPFC stock
assessment (Davies, et al. 2014) are convenient potential abundance
indices. The assessment method used by the
WCPFC is spatially structured and estimates biomass in several regions
in the Pacific Ocean. The boundary between regions 2 and 4 passes
directly through the MHI (Figure~\ref{fig:mhimap}).
Biomass estimates from both regions were evaluated as abundance indices.

The logistic parameters $r$ and $K$ are notoriously difficult to estimate
in surplus production models, 
so parameter substitutions were used. 
Maximum sustainable yield
(MSY, $\MSY$) and fishing mortality at MSY ($\Fmsy$) are commonly used
fishery reference points and are simple functions of $r$ and $K$.
Therefore $\MSY$ and $\Fmsy$ were
estimated directly and substituted in (\ref{eqn:intschaeferA}) as
$r=2\Fmsy$ and $K=\frac{4\MSY}{r}$.

Carruthers and McAllister (2011) recommend use of Bayesian priors for the
logistic growth parameter, $r$, in equation~(\ref{eqn:ischaefer}). They
suggest $\tilde{r} = 0.486$ with a standard deviation of $\sigma_r = 0.046$
(a coefficient of variation ${\rm cv}=\frac{\sigma_r}{r}=0.094$)
for Atlantic YFT.
A lognormal prior on $r$ was implemented as 
\begin{equation}
\log r - \log \tilde{r} = \rho ;\quad \rho\sim N(0,\sigma^2_r);
\end{equation}
$\rho$ becomes a component of the likelihood
(equation~\ref{eqn:likelihood}).

This model assumes three Normal $N(0,\sigma^2)$ process errors with
standard deviations, $\sigma_N$, $\sigma_I$, and $\sigma_F$. The first
two, $\sigma_N$ and $\sigma_I$, pertain to the evolution of biomass over
time. It is assumed that these two standard deviations can be
represented by a single parameter, $\sigma_P$. The third standard
deviation, $\sigma_F$, pertains to the fishing mortality.
\pagebreak[4]

The general form of the state space {\bf observation model} is
\begin{equation}
x_t = O(\alpha_t) + \varepsilon_t
\end{equation}
where the function $O$ describes the measurement process with
error $\varepsilon$ in observing the state.

Predicted catch, $\widehat{C}_{g,t}$, for each gear is the product of
estimated fishing mortality and the biomass,
\begin{equation}
\widehat{C}_{g,t} = F_{g,t}\cdot\Bigl(\frac{N_{t-\Delta
t}+N_t}{2}\Bigr) \cdot e^{\varepsilon_t}
\label{eqn:obs1}
\end{equation}
where the biomass is  the average
biomass over the time step (Quinn and Deriso, 1999), and
$\varepsilon_t$ is a ``zero-inflated'' log normal likelihood given by
\begin{equation}
  \log \varepsilon_t = \left\{
    \begin{array}{r@{\;:\quad}l}
       C_{g,t} > 0 &
(1-p_0)\cdot\bigg(\log\frac{1}{\sqrt{2\pi\sigma^2_Y}}
          -\Bigl(\frac{\log
C_{g,t}-\log\widehat{C}_{g,t}}{\sigma_Y}\Bigr)^2\bigg)\\
       C_{g,t} = 0 & p_0 \cdot\log \frac{1}{\sqrt{2\pi\sigma^2_Y}}\\
    \end{array}
  \right.
\end{equation}
where $C_{g,t}$ is the observed catch for gear $g$ at time $t$,
$\sigma_Y$ is the observation error and
$p_0$ is the proportion of observed catch observations equal to zero.

\begin{table}
\caption{Complete list of estimated and computed parameters and model
constraints for the state-space surplus production model. 
There are 6 estimated parameters. 
The three computed variables are functions of estimated parameters.
There are four constraints and constants.
For a model with 61 time steps and 4 gear types, there are 305 random effects.
}
\label{tab:allvars1}
\begin{center}
\begin{tabular}{ll}
\hline
Parameter & Definition\\
\hline
\hline
       & {\it Estimated parameters:}\\
$\MSY$ & Maximum sustainable yield (mt).\\
$\Fmsy$& Fishing mortality at maximum sustainable yield (\peryr).\\
$Q$    & Abundance index proportionality constant\\
$\sigma_P$ & Biomass random walk process error standard deviation,\\
           & $\sigma_P=\sigma_N=\sigma_I$ (mt).\\
$\sigma_F$ & Fishing mortality random walk process error\\
           & standard deviation (\peryr).\\
$\sigma_Y$ & Observation error standard deviation (mt).\\
\hline
       & {\it Computed variables:}\\
$r$    & Instantaneous growth rate, $r=2F_{\MSY}$ (\peryr).\\
$K$    & Asymptotic population size, $K=\frac{4\MSY}{r}$ (mt).\\
$\bar{F}_5$ & Average of total estimated fishing mortality\\
            & for the most recent five years (\peryr).\\
\hline
       & {\it Constraints and constants:}\\
$p_0$  & Proportion of zero catch observations,\\
       & fixed at 0.04918\\
$\tilde{r}$ & An {\it a priori} assumed value for $r$ (\peryr)\\
            & fixed at $\tilde{r}=0.486$ (Carruthers and McAllister, 2011)\\
$\sigma_r$  & Assumed standard deviation of $r$ around its prior,\\
            & fixed at $\sigma_r=0.8$ or $\sigma_r=0.1$ (\peryr).\\
$\bar{Y}_5$ & Average of total observed catch\\
            & for the most recent five years (818 mt).\\
\hline
\end{tabular}
\end{center}
\end{table}

The model states, $N_t$ and $F_{gt}$, are assumed to be random
effects (Skaug and Fournier, 2006). Model parameters are estimated by
maximizing the joint likelihood of the random
effects, observations, and prior likelihood.
\begin{equation}
\label{eqn:likelihood}
L(\theta,\alpha,x)=
\prod^m_{t=2}\big[\phi\big(\alpha_t-T(\alpha_{t-1}), \Sigma_\eta\big)\big]\cdot
\prod^m_{t=1}\big[\phi\big(x_t-O(\alpha_t),
\Sigma_\varepsilon\big)\big]\cdot\rho
\end{equation}
where $m$ is the number of time steps in the catch time series and
$\theta=\left\{\MSY,\Fmsy,Q,\sigma_P,\sigma_F,\sigma_Y\right\}$ 
is a vector of model parameters (Table~\ref{tab:allvars1}).
The model is implemented in ADMB-RE (Fournier et al. 2012) and
applied to a 61 year ($m=61$) time series of reported commercial
catch by four fishing gears from 1952 through 2012
(Figure~\ref{fig:data}).
Data preparation is described in Appendix~\ref{sec:data}.
Noncommercial fishers land substantial quantities of YFT, but
reliable data from this sector of the fishery are not currently
available. The effects of omitting the noncommercial catch is
discussed briefly in Appendix~\ref{sec:klingon}.
All computer code and data files discussed in this
paper can be found at Github:
\url{https://github.com/johnrsibert/XSSA.git}.

\section*{Results}
Table~\ref{tab:ests1} compares 6 different models
using three index of abundance assumptions and two 
assumed values of the standard deviation of the $r$ prior, $\sigma_r$. 
Models without an index of abundance do not converge to solutions as
indicated by the values of $|G|_{max}$.
Values of $-\log L$ are lower for models with the loose constraint on $r$
($\sigma_r=0.8$) than for equivalent models with the tight constraint on $r$
($\sigma_r=0.1$).
Models with $\sigma_r=0.1$ estimate higher values of $\Fmsy$ and the
calculated values of $r$ are, of course, closer to the prior.

\begin{table}
{\small
\caption{Results for 6 different models using three
abundance indices and two constraints on $r$. The first section of the
table shows model diagnostics. $-\log L$ is the negative logarithm of
equation~\ref{eqn:likelihood}; lower values indicate better fits to
the data.
$|G|_{max}$ is the maximum gradient of the likelihood with respect to
the parameters; values of $|G|_{max}>10^{-4}$ (1e-04) indicate lack of
model convergence.
$n$ is the number of parameters estimated.
All values are reported to three significant figures.
}
\label{tab:ests1}
\begin{center}
\begin{tabular}{|l|rrr|rrr|}
\hline
Index Region  &  2& 4& None& 2& 4& None\\
\hline
$-\log L$ & 157.54 & 128.73 & 789.98 & 170.78 & 130.36 & 317.17\\
$n$ & 6 & 6 & 5 & 6 & 6 & 5\\
$|G|_{max}$ & 9.72e-06 & 8.65e-06 & 0.266 & 3.5e-05 & 5.44e-06 & 0.455\\
\hline
$\MSY$ & 1180 & 765 & 3040 & 1030 & 928 & 1490\\
$\Fmsy$ & 0.0235 & 0.0815 & 0.248 & 0.207 & 0.225 & 0.239\\
$Q$ & 0.438 & 0.00699 & --- & 0.025 & 0.00288 & ---\\
$\sigma_P$ & 0.112 & 0.0699 & 0.00203 & 0.167 & 0.0789 & 0.00366\\
$\sigma_F$ & 0.323 & 0.346 & 0.328 & 0.321 & 0.345 & 0.336\\
$\sigma_Y$ & 0.393 & 0.382 & 0.389 & 0.394 & 0.382 & 0.393\\
\hline
$r$ & 0.047 & 0.163 & 0.495 & 0.415 & 0.45 & 0.478\\
$K$ & 100000 & 18800 & 24500 & 9970 & 8240 & 12500\\
$\bar{F}_5$ & 0.0104 & 0.181 & 0.388 & 0.19 & 0.429 & 0.334\\
\hline
$p_0$ & 0.197 & 0.197 & 0.197 & 0.197 & 0.197 & 0.197\\
$\tilde{r}$ & 0.486 & 0.486 & 0.486 & 0.486 & 0.486 & 0.486\\
$\sigma_r$ & 0.8 & 0.8 & 0.8 & 0.1 & 0.1 & 0.1\\
\hline
\end{tabular}
\end{center}
}
\end{table}

\begin{figure}
\begin{center}
\includegraphics[height=0.45\textheight]{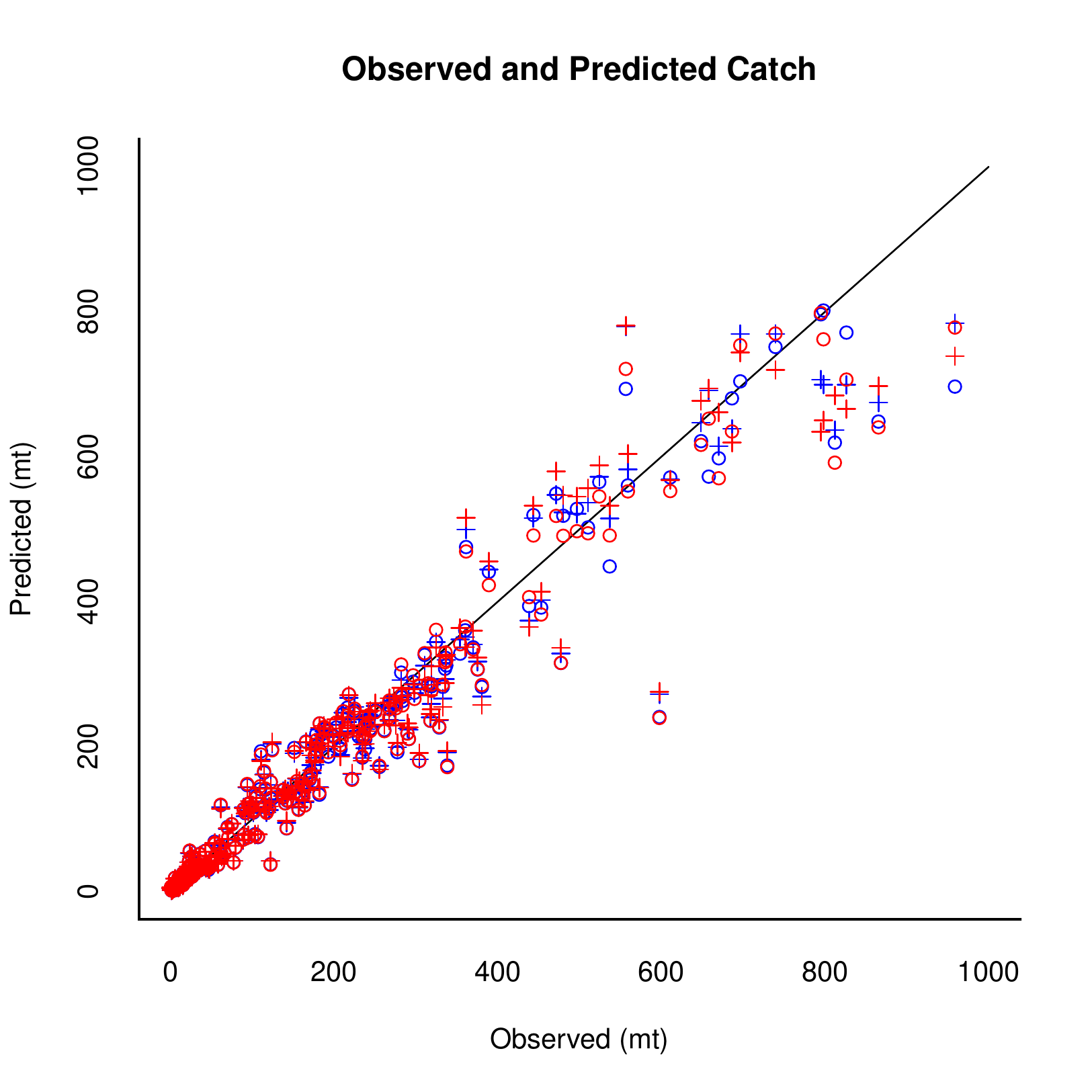}
\caption{Plot of predicted catch on observed catch for two indices of
abundance.
Open circles (o) indicate region 2 index; plus (+) symbols indicate
region 4 index.
Blue indicates $\sigma_r=0.8$; red indicates $\sigma_r=0.1$.
The black line has a slope of 1 and passes through the origin.
\label{fig:mopcatch}
}
\end{center}
\end{figure}

\pagebreak[4]
Models with region 2 indexing and region 4 indexing appear to be
equally capable of accurately predicting catch,
Figure~\ref{fig:mopcatch}. Low catches are predicted
accurately, but some high catches are estimated with larger errors, as is
expected with log-normal errors. 
See Figure~\ref{fig:estC} for more detail.

\begin{figure}
\begin{center}
{\scriptsize \sffamily
\begin{tabular}{cc}
A. Region 2 index; $\sigma_r=0.8$ &
B. Region 4 index; $\sigma_r=0.8$ \\
\includegraphics[height=0.35\textheight]{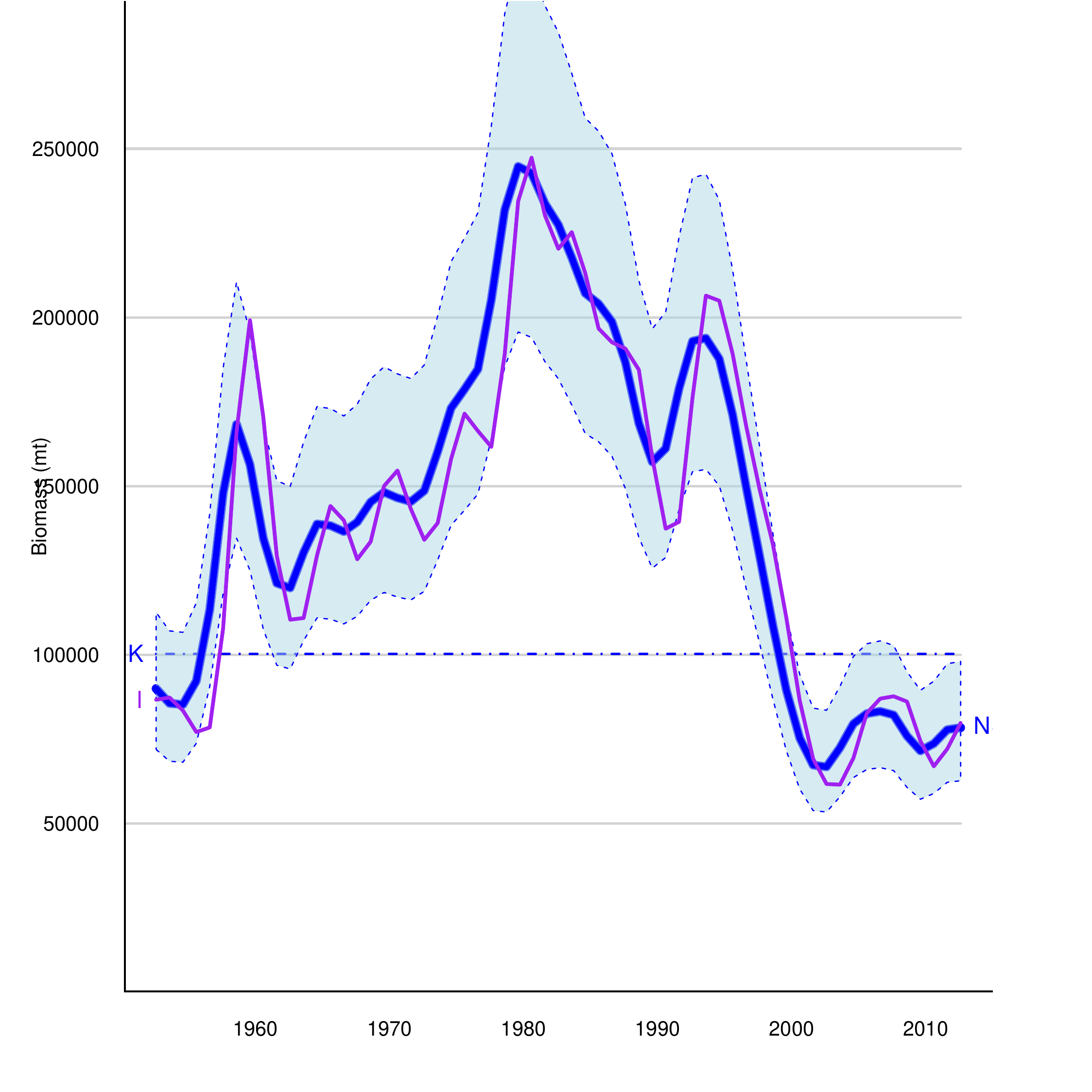} &
\includegraphics[height=0.35\textheight]{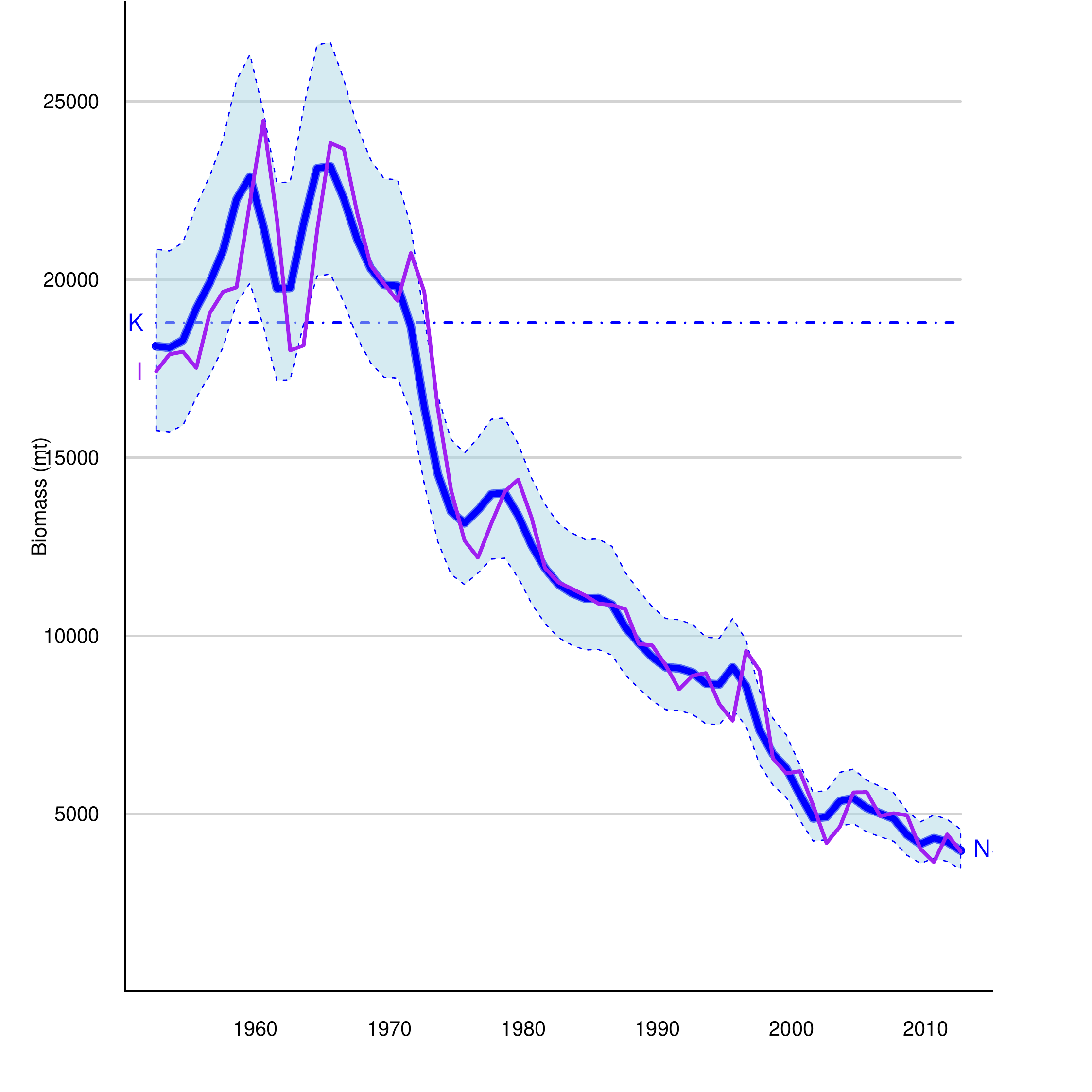} \\
C. Region 2 index; $\sigma_r=0.1$ &
D. Region 4 index; $\sigma_r=0.1$ \\
\includegraphics[height=0.35\textheight]{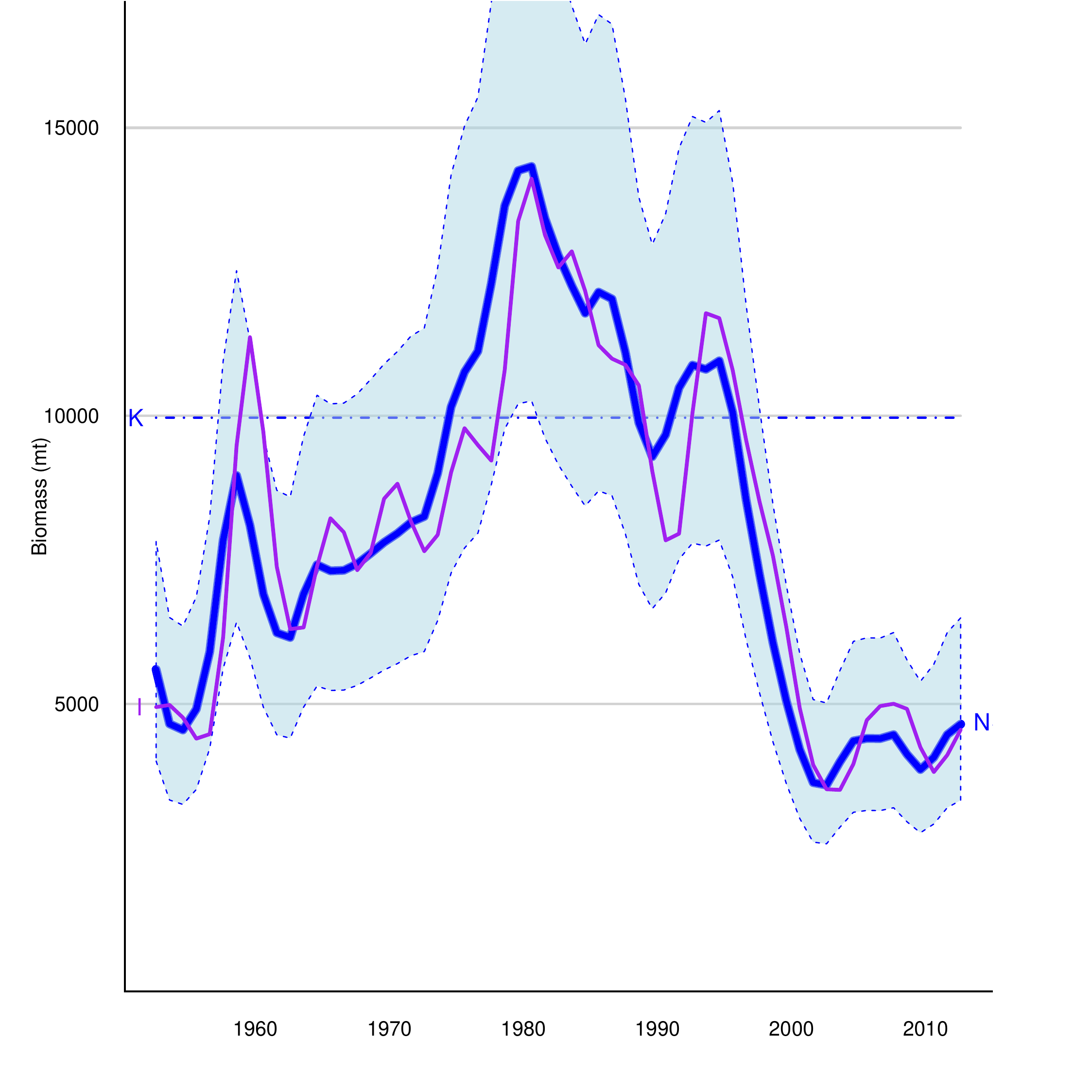} &
\includegraphics[height=0.35\textheight]{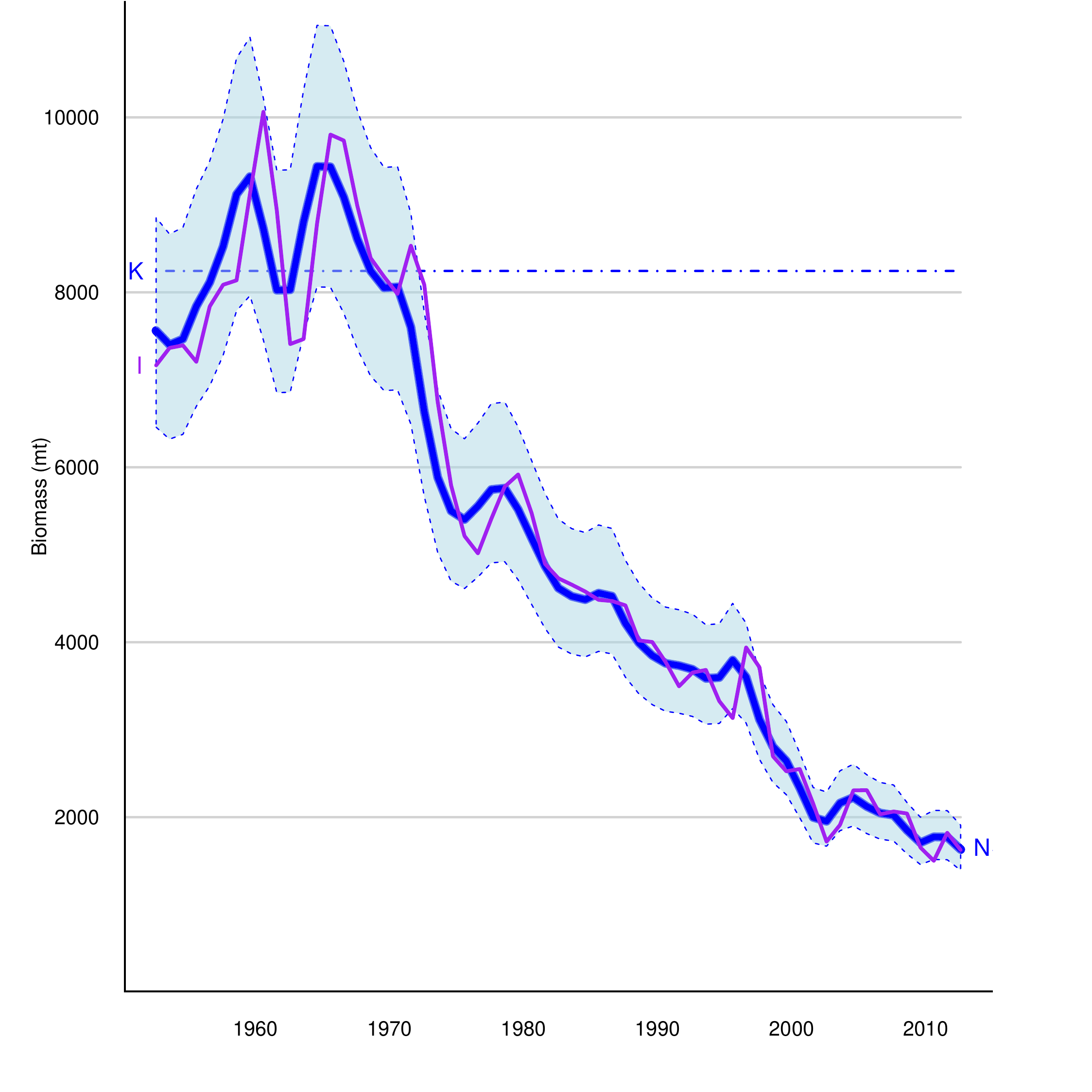} \\
\end{tabular}
}
\caption{Predicted biomass trends with different indices of abundance
and values of $\sigma_r$.
Blue lines indicate the estimated biomass in metric tons (mt);
purple lines indicates the index of abundance.
The light blue shaded areas represent the process error as 
$\pm 2\sigma_P$.
The equilibrium biomass $K$ is indicated by the blue dot-dash line.
Note that the scale of the ordinate is different in each panel.
\label{fig:estbiomass}}
\end{center}
\end{figure}

Estimated biomass tracks abundance indices for all models,
Figure~\ref{fig:estbiomass}, and the abundance indices lie within the
process error of the biomass trend.
The maxima of the estimated biomass trends may exceed the
estimated equilibrium biomass ($K$) for varying periods of time for
all models.
In the case of the model with region 2 index $\sigma_r=0.8$,
the estimated biomass exceeds $K$
for most of the time series.

The average total YFT catch in the MHI for the most recent 5 years
is 818 mt. The estimated $\MSY$ is higher than this average for all
models except the region 4 index with $\sigma_r=0.1$.
The estimated average fishing mortality for the last 5 years is below
the estimated $\Fmsy$ for both region 2 models, but above estimated
$\Fmsy$ for the region 4 models; see Table~\ref{tab:ests1}.

\section*{Discussion}
Failure of the model to converge to a solution without an index of
abundance (Table~ \ref{tab:ests1})
indicates that the data lack sufficient information for this model.
An informative prior constraint on a parameter and an index of abundance are 
additional assumptions that provide information to a model
to constrain its behavior.
Interpretations of these two assumptions are quite different. 
Assuming a prior value for $r$ implies {\it a priori} knowledge of stock
productivity, and imposing a low standard deviation around the
prior implies a high level of certainty about its value.
On the other hand,
assuming that local stock size is proportional to stock size in
a larger geographic range
implies that factors mediating the abundance
of the local stock are similar to those mediating the
abundance in the larger area.

MFCL Regions 2 and 4 differ from one another in their oceanography,
ecology and fisheries (Appendix~\ref{sec:index}).
Region 2 is outside the center of abundance for YFT in the
WCPO, appears to have a relatively small YFT stock, and Davies et al. (2014)
estimate that region 2 has experienced relatively minor fishery impact.
In contrast, region 4 straddles the equatorial center of YFT abundance,
appears to have a larger stock, hosts some of the largest fisheries in the WCPO,
produces a large proportion of the total WCPO yellowfin catch, and
Davies et al. (2014) estimate that region 4 has experienced
the largest impact of the fishery of any of the MFCL regions.
Region 2, therefore, is the best {\it a priori} choice for
an index of abundance.

The random walk representation of fishing mortality enables
the model to compute accurate values of $F_{g,t}$. The zero-inflated
log normal observation error and the separation of fishing fleets
enable the model to interpret low (or zero) catches as changes in
$F_{g,t}$ rather than as episodes of low stock size.
These features enable the model to estimate catch
from the estimated stock extremely accurately.

Current catches ($\bar{Y}_5=818$) are less than model estimates of maximum
sustainable yield ($\MSY$) for all model variants. 
Estimates of current fishing mortality ($\bar{F}_5$) are less than
estimates of fishing mortality at maximum sustainable yield ($\Fmsy$)
for all models with region 2 index of abundance.
If fishery managers were to adopt $\MSY$ and $\Fmsy$ as management
reference points, the conclusion would be that the local stock is not
overfished and that overfishing is not occurring.

The model presented here is a promising, if preliminary,
approach to assessing a local
portion of a larger fish stock in a way that provides useful advice to
local fisheries managers. The current model is clearly in need of
further development. Estimates of noncommercial catch (and possibly
their observation error) should be included in the data. 
The issue of selecting an appropriate index of abundance should be
resolved. One hopes that the
next round of  WCPFC stock assessments will realign
the MFCL regions to be more consistent
with the Longhurst (1998) ecological provinces so that the entire
Hawaiian Archipelago lies within a single stock assessmemt region. 
A region that aligns approximately with  the North Pacific Tropical
Gyre Province would be the best choice for an index of abundance for
the MHI.


\clearpage
\singlespacing
\noindent {\bf Acknowledgements.}
This work was initiated under a contract from the Western Pacific
Regional Fisheries Managment Council. 
I thank the Council for its generous support and
Council Staff Paul Dalzell and Eric Kingma for encouraging me to
take on this challenging project.
Special thanks are due 
to David Itano for sharing insights into the biology of yellowfin tuna
and into small-boat fisheries in Hawaii,
to Reginald Kokubun of the State of Hawaii, Division of Aquatic Resources,
Department of Land and Natural Resources, for supplying catch report
data from the Commercial Marine Landings data base,
to Keith Bigelow and Karen Sender of NOAA Pacific
Island Fisheries Science Center for supplying logbook reporting data and
from the PIFSC data base.
Thanks also to John Hampton of the Secretariat of the Pacific
Community, Oceanic Fisheries Programme, for making available \MFCL\
output files from the latest Western and Central Pacific
Fisheries Commission yellowfin tuna stock assessment, and to Nick
Davies for advice on interpreting those files.

\section*{References}
{\parindent=0cm \small
\everypar={\hangindent=2em \hangafter=1}\par
Carruthers, T. and M. McAllister. 2011.
Computing prior probability distributions for the
intrinsic rate of increase for Atlantic tuna and
billfish using demographic methods.
Collect. Vol. Sci. Pap. ICCAT, 66(5): 2202-2205.

Davies, N., S. Harley, J. Hampton, S. McKechnie. 2014. Stock
assessment of yellowfin tuna in the western and central pacific ocean.
WCPFC-SC10-2014/SA-WP-04.

Fournier, D. A., H.J. Skaug, J. Ancheta, J.Sibert, J. Ianelli, 
A. Magnusson, M. N. Maunder, A. Nielsen. 2012. AD Model Builder:
using automatic differentiation for for statistical inference of highly
parameterized complex nonlinear models. Optimization Methods and
Software 27, 233–249.

DLNR. 2011. HMRFS Newsletter. Hawaii Department of Land and Natural
Resources.https://dlnr.hawaii.gov/dar/fishing/hmrfs/

Longhurst, A. 1998. {\it Ecological Geography of the Sea}. Academic
Press, San Diego. 398pp.

Nielsen, A., C. Berg. 2014. Estimation of time-varying selectivity
in stock assessments using state-space models. Fisheries Research
158:96-101.

Quinn, T, R. Deriso. 1999. Quantitative fish dynamics. Oxford
University Press, New York.

Rooker, J.R., R. J. D. Wells, D. G. Itano, S. R. Thorrold, J. M. Lee. 2016.
Natal origin and population connectivity of bigeye and
yellowfin tuna in the Pacific Ocean.
Fish. Oceanogr. 25, 277-291.

Schaefer, M. B. 1954. Some aspects of the dynamics of populations
important to the management of the commercial marine fisheries. IATTC
Bull. 1:27-56.

Senina, I.,  P. Lehodey, B. Calmettes, S. Nicol, S. Caillot,
J. Hampton and P. Williams. 2015.
SEAPODYM application for yellow tuna in the Pacific Ocean.
WCPFC-SC11-2015/EB-IP-01.

Skaug, H., Fournier, D., 2006. Automatic approximation of the marginal
likelihood in non-Gaussian hierarchical models. Computational
Statistics \& Data Analysis 51, 699–709.


Wilson, P. T. 2011. {\it AKU!}. Xlibris, USA. 368pp. ISBN
978-1-4568-5904-6.
\par}

\clearpage

\numberwithin{figure}{section}
\appendix
\section{Data preparation}
\label{sec:data}

\begin{figure}
\begin{center}
\includegraphics[height=0.8\textheight]{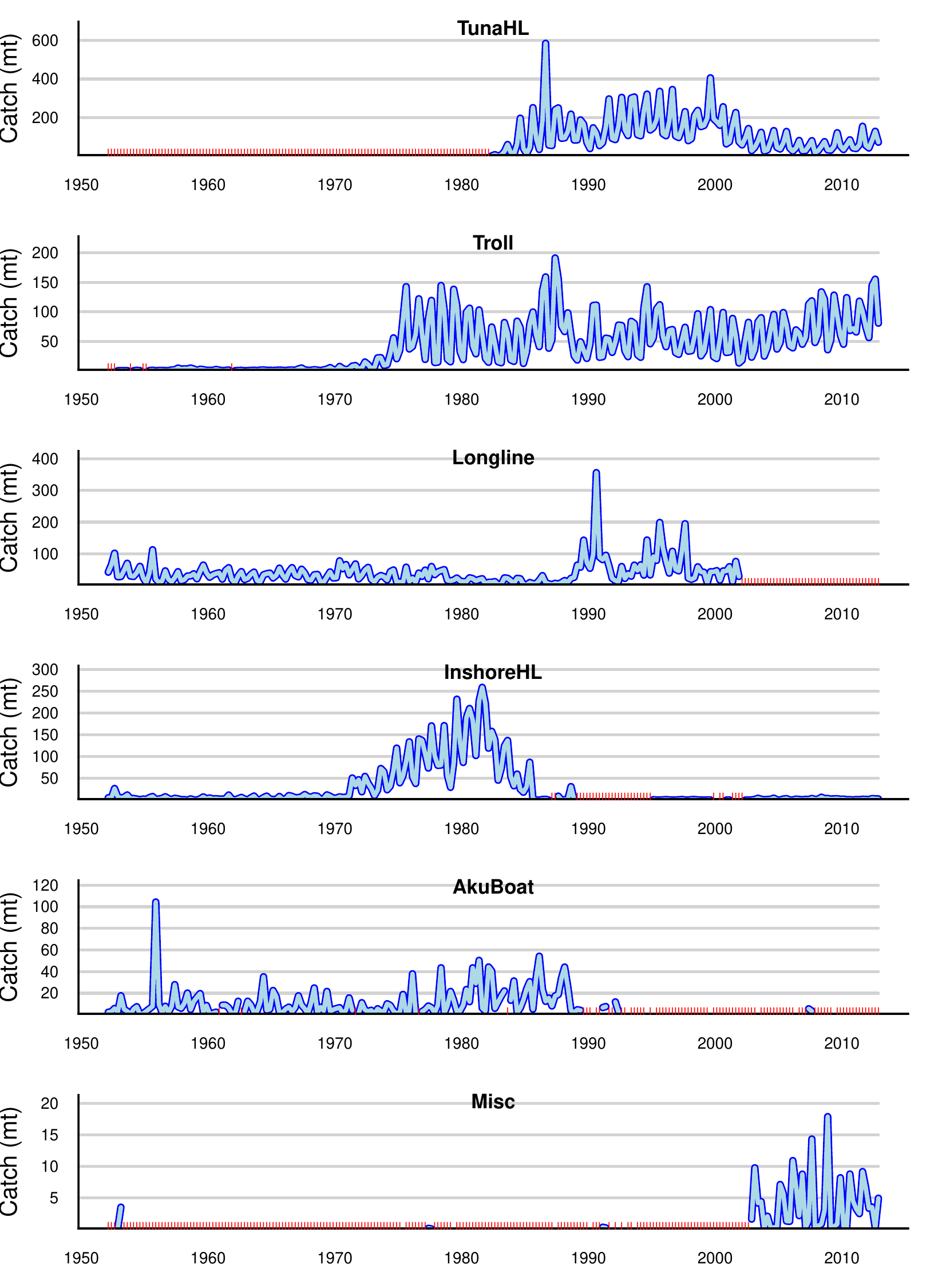}
\caption{\label{fig:CMLdata}
Yellowfin catch in metric tonnes by principle fisheries operating in
the Main Hawaiian Islands from the State of Hawaii Commercial Marine
Landings Data.
The red tick marks on the abscissa indicate quarters where reported
catches were zero.
}
\end{center}
\end{figure}

The State of Hawaii Department of Land and Natural Resources
Commercial Marine Landings data base (CML) from 1949 through 2014 is
the primary source of data used in this analysis. This database
documents the
continuous 65 year history of commercial fishing in Hawaii.

The CML catch reports were aggregated into the following gear categories:
``Aku boat'', ``Bottom/inshore HL'', ``Longline'',  ``Troll'', ``Tuna
HL'', ``Casting'', ``Hybrid'',  ``Shortline'', ``Other'', and
``Vertical line''.
For this analysis catches by ``Casting'', ``Hybrid'',
``Shortline'', ``Other'', and ``Vertical line'' are combined into a new
category, ``Misc''. Landing in the ``Misc'' category are highest after
year 2000 and less than  2\% of the total landings.
The catch time series for the CML data are shown in
Figure~\ref{fig:CMLdata}.
The ``Bottom/inshore HL'' and ``Tuna HL'' are both handline gears and
the data for these two gear types were combined at the suggestion of
the CML database administrator.
The ``Aku Boat'' gear type refers to storied Japanese-style pole and
line fishery that operated in Hawaii through th 1980s (Wilson 2011). 
The Aku Boats target skipjack tuna 
({\it Katsuwonus pelamis}), but yellowfin were occasionally landed as
incidental catch.
Some time series contain sustained periods of zero catch which
reflect the development and subsequent shift away from a specific
gear type. It is assumed that these declines in catches represent
``collapse'' of a fishery due to social and economic factors rather than
to a decline of YFT stocks.
Similarly, some time series are punctuated by brief episodes (one or two quarters in
length) of zero catches. Again, it is assumed that these zero catches
are not caused by low stock size.

\begin{figure}
\begin{center}
\includegraphics[height=0.8\textheight]{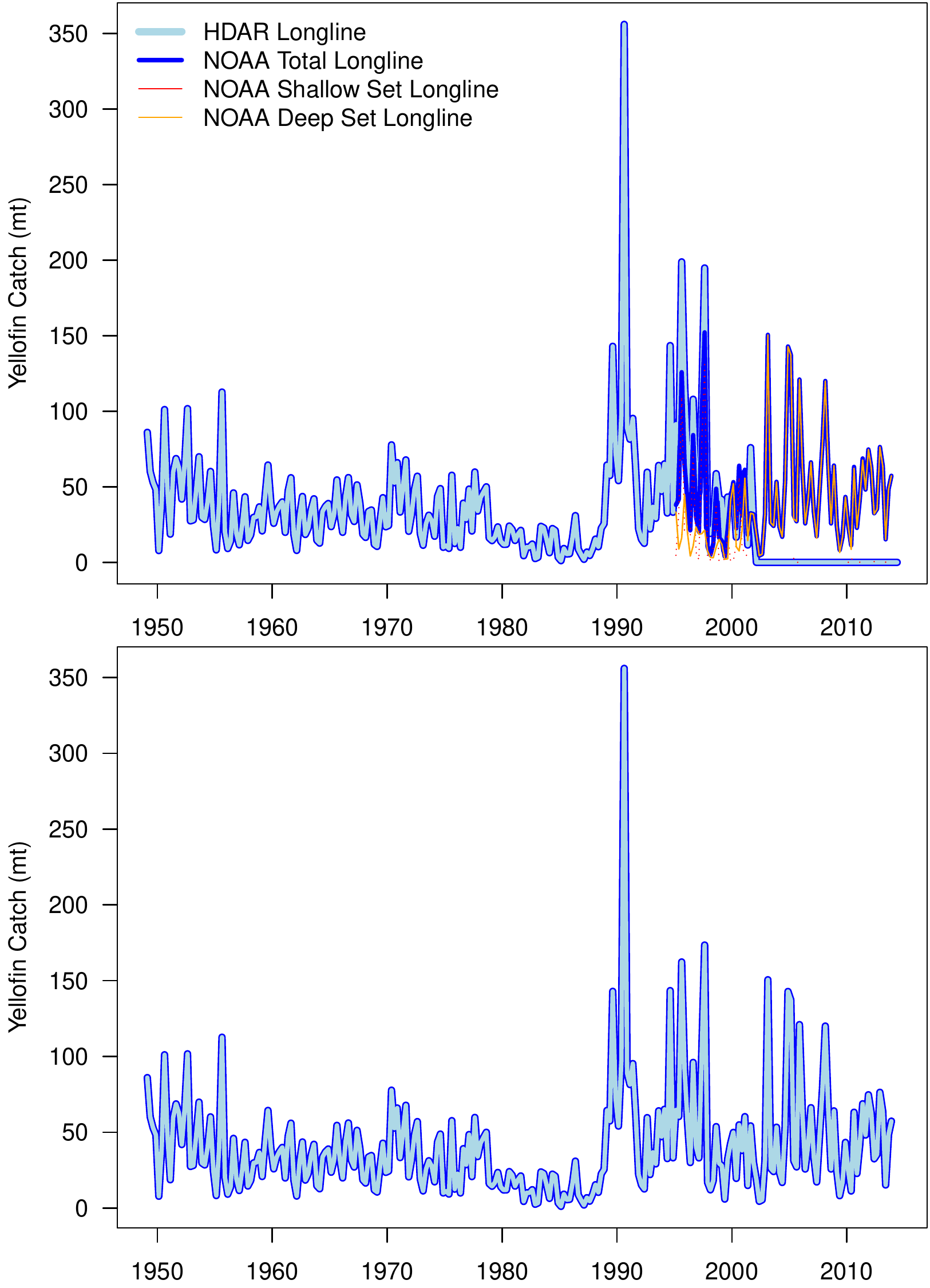}
\caption{\label{fig:hdarnoaaLLTS}
Comparison between CML and NOAA longline time series. The upper panel
shows the NOAA deep and shallow set data superimposed on the HDAR
data. The lower panel shows the time series produced by a simple
average of the CML data and the sum of the NOAA deep and shallow
catches.}
\end{center}
\end{figure}

The Hawaii-based longline fishery began a rapid expansion in the late
1980s, and the United States National Oceanic and Atmospheric Administration (NOAA)
began to collect data from the longline fleet under a
federally mandated logbook program in 1990. The CML longline data were
augmented by NOAA longline log sheet data from 1995 through 2013.
NOAA distinguishes deep sets, targeting bigeye tuna ({\it Thunnus
obesus}), and shallow sets,
targeting swordfish ({\it Xiphias gladius}) in the data. 
The CML data do not distinguish between deep and shallow sets.
Since the longline fleet ranges widely in the North Pacific Ocean,
only catches reported in the United States EEZ around Hawaii east of
162\deg{W} longitude were included in the data.
Figure~\ref{fig:hdarnoaaLLTS} shows the correspondence between the
CML and NOAA time series. The combined deep plus shallow catches from
NOAA align fairly well with the overlapping CML data. The simple
average of the CML data with the combined NOAA deep plus shallow data
appears to be roughly the same trajectory as the
constituent time series. Yellowfin is currently considered an incidental catch
in the longline fishery.

\begin{figure}
\begin{center}
\includegraphics[height=0.8\textheight]{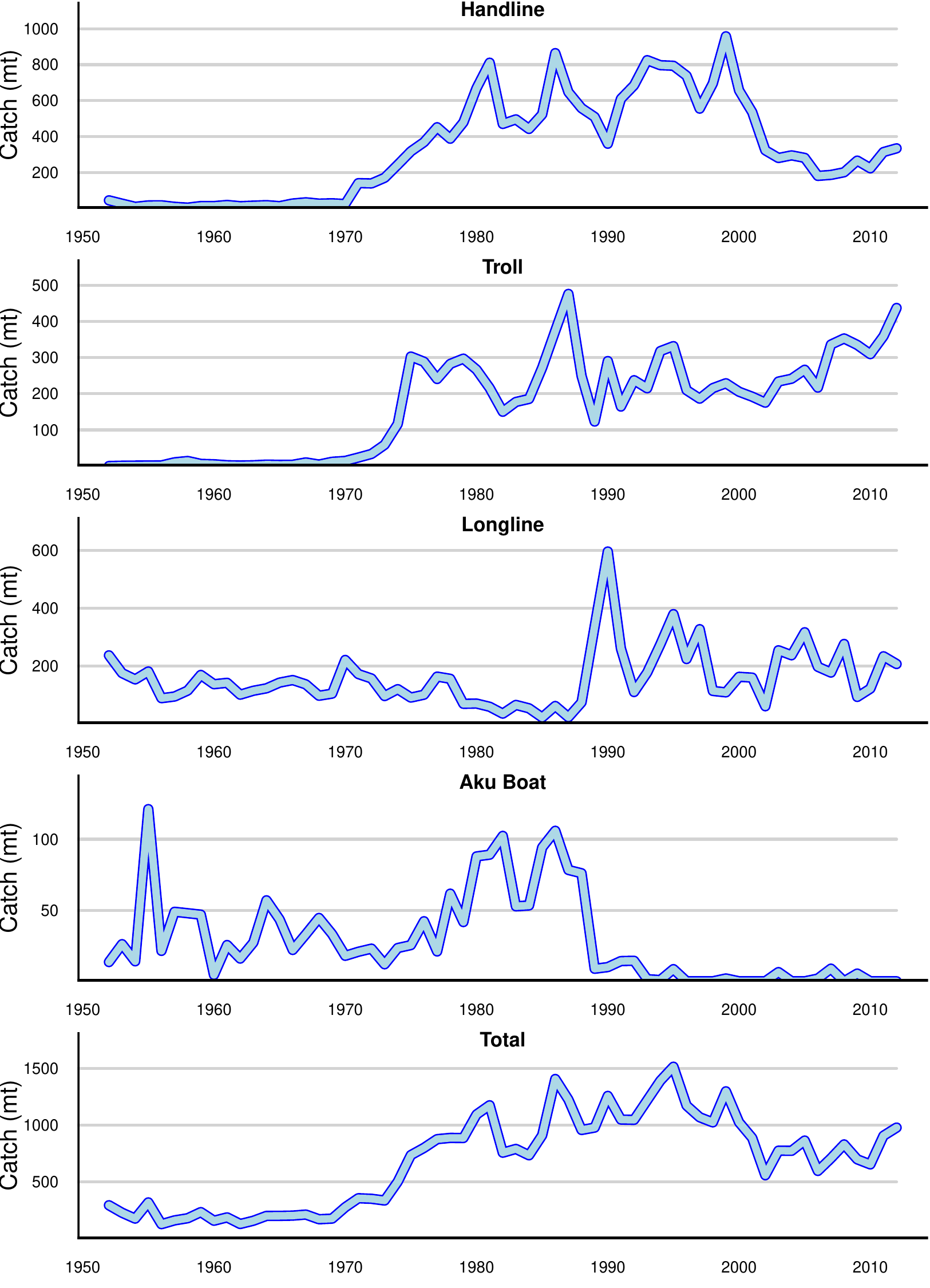}
\caption{\label{fig:data}
Four gear type catch time series used in the assessment model.
}
\end{center}
\end{figure}

Catch data from the CML and NOAA longline data bases were initially
aggregated by quarter of the calendar year.
All catch time series exhibit annual cycles suggesting strong seasonal
signals in the catches by all gears. To avoid the need to estimate
autocorrelation matrices for each time series and to minimize the
number of zero catch observations, the quarterly time series were
aggregated into the annual time series shown in Figure~\ref{fig:data}
and used in this analysis.

\clearpage
\section{Abundance indices and offline coupling}
\label{sec:index}
\begin{figure}
\begin{center}
\includegraphics[width=0.9\textwidth]{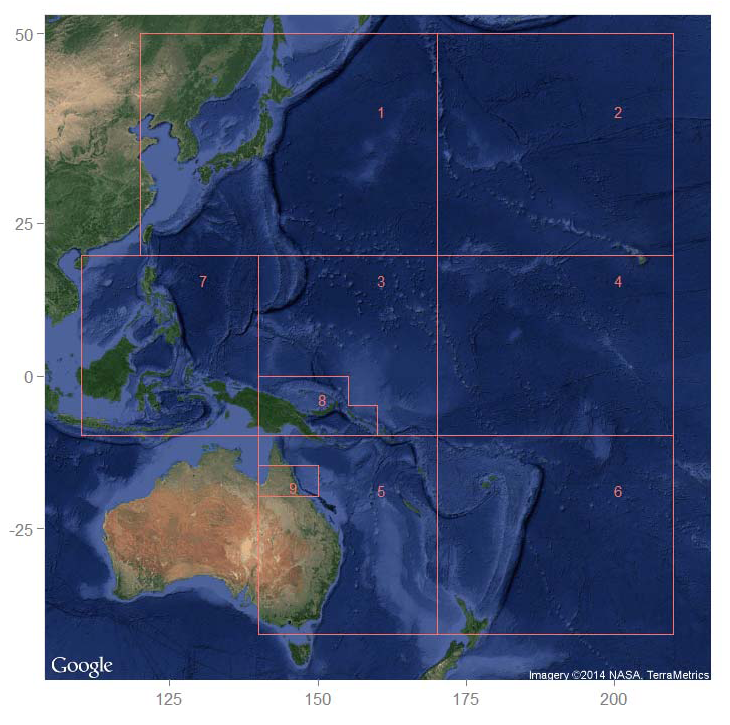}
\caption{\label{fig:mfclregions}
Regional structure used in the 2014 WCPFC YFT stocks assessment from
Davies et al. (2014).
}
\end{center}
\end{figure}

Using the output of one model to constrain a second model is sometimes
referred to as ``offline'' coupling. 
The output from two quite different models of WCPO yellowfin,
\SD\ and \MFCL, could potentially be used as indices of abundance.
Both models are age structured, spatially structured, and have been
applied to YFT.
\SD\ is an ocean basin scale model with 1\deg{} spatial resolution,
whereas the \MFCL\ yellowfin assessment model is constrained to the
western Pacific and partitions the model domain into 9 regions,
Figure~\ref{fig:mfclregions}.
The spatial resolution of the \SD\ model would make it an ideal
candidate for computing an index of abundance, but the model and its
application to yellowfin are still under active development (Senina et
al, 2015). 
The 2014 \MFCL\ assessment (Davies et al. 2014) has been officially
adopted by the WCPFC as the assessment on which regional conservation
and management measures are based.

\begin{figure}
\begin{center}
\includegraphics[width=1.10\textwidth]{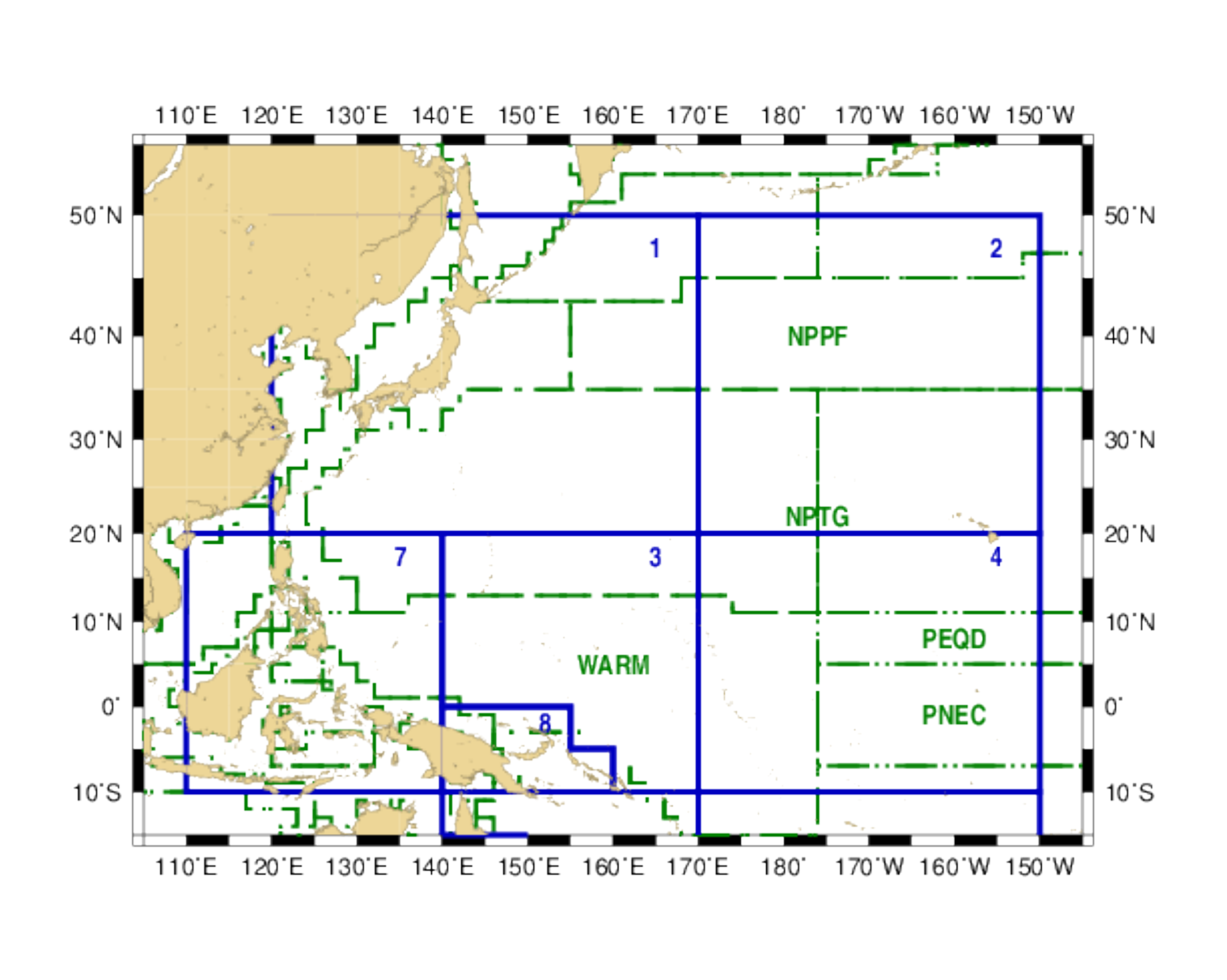}
\caption{Map of the central north Pacific Ocean showing selected \MFCL\
stock assessment regions (blue lines and numbers) and 
Longhurst (1998) ecological provinces (dashed green lines and labels);
NPFF, North Pacific Transition Zone Province;
NPTG, North Pacific Tropical Gyre Province;
WARM, Western Pacific Warm Pool Province;
PEQD, Pacific Equatorial Divergence Province;
PNEC, North Pacific Equatorial Countercurrent Province.
\label{fig:paczones}}
\end{center}
\end{figure}

\begin{figure}
\begin{center}
{\scriptsize \sffamily
\begin{tabular}{cc}
MFCL Region 2 & MFCL Region 4\\
\includegraphics[width=0.50\textwidth]{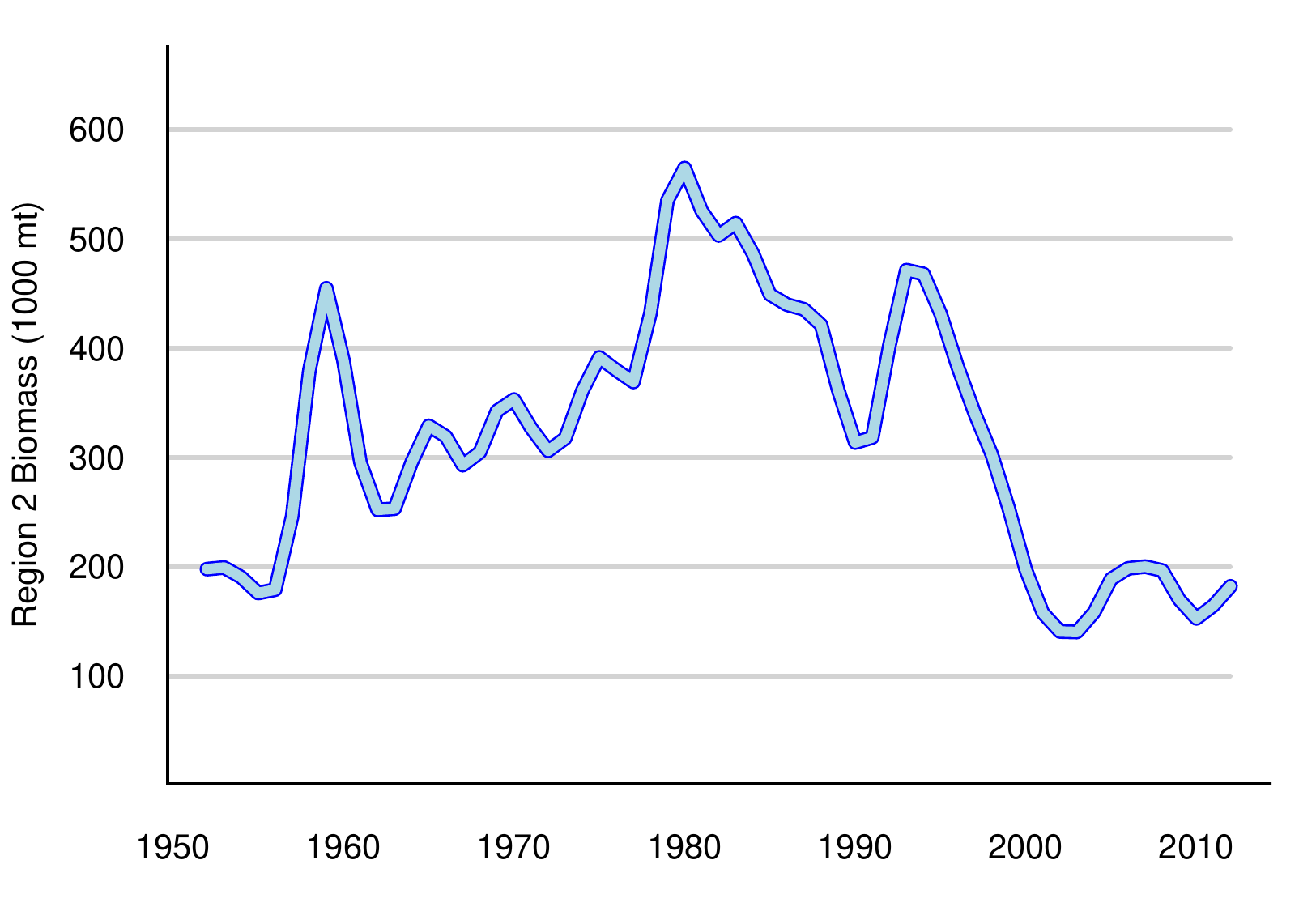} &
\includegraphics[width=0.50\textwidth]{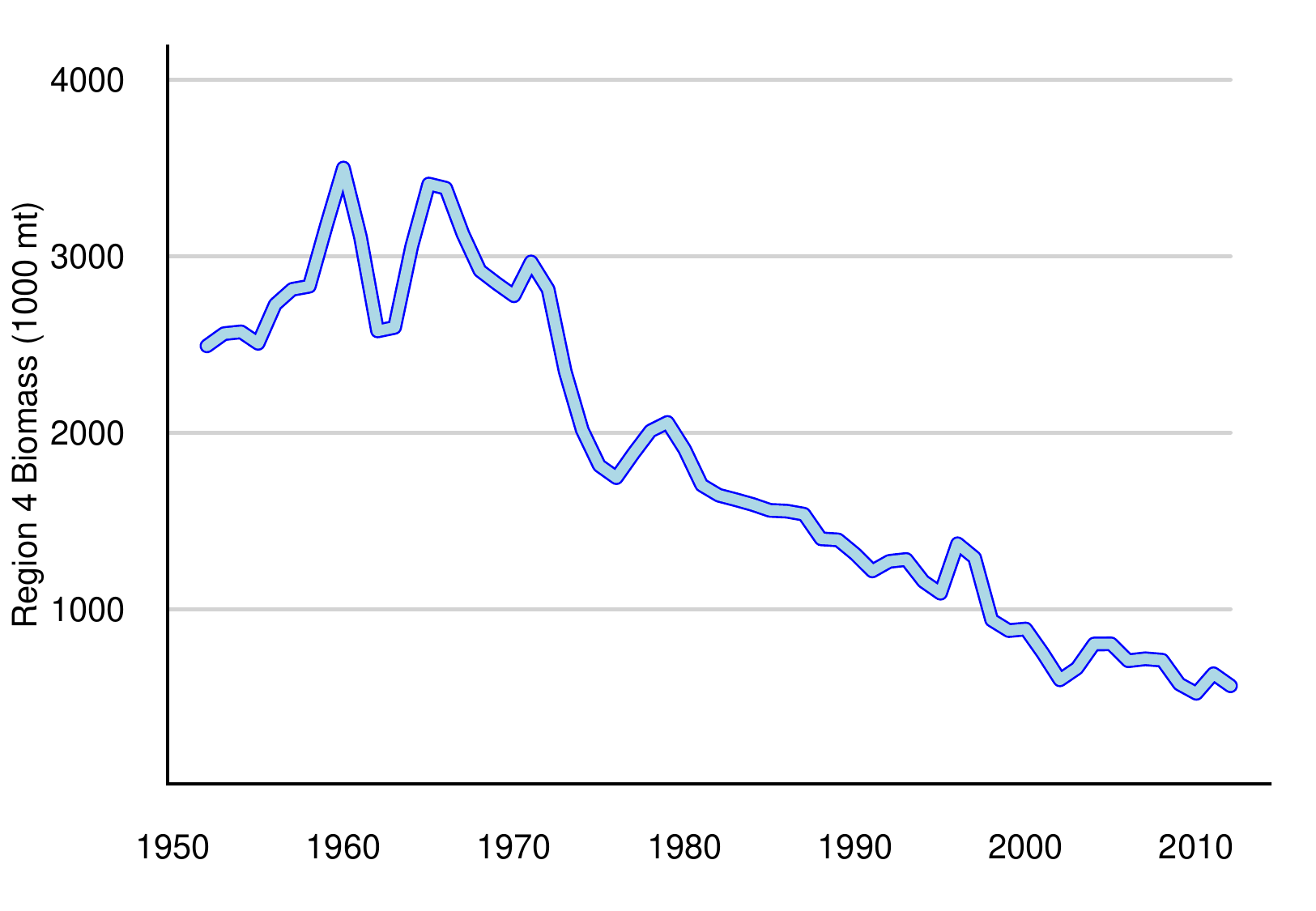} \\
\end{tabular}
}
\end{center}
\caption{Estimated biomass trends in \MFCL\ regions 2 and 4 (Davies, et al. 2014).
\label{fig:mfclbiomass}
}
\end{figure}

The Hawaiian archipelago lies in the eastern extension of Longhurst's
(1998) North Pacific Tropical Gyre Province (NPTG), 
Figure~\ref{fig:paczones}.
The boundary between MFCL regions 2 and 4 splits both MHI and the NPTG.
Region 2 includes the northern extent of the range of YFT. The MFCL
assessment concludes that the YFT biomass is relatively low and that
the impact of the WCPO fisheries has been relatively minor.
In contrast, region 4 lies in primarily in the Western Pacific Warm
Pool Province, includes the core of the YFT range in the Pacific,
and supports some of the most intense tuna fisheries in the world.
MFCL estimates that the impact of the fisheries on the YFT stock in
region 4 to be one among the highest of all MFCL regions, and it is
likely that overfishing is occuring.
The estimated biomass trends in these two regions are quite different,
Figure~\ref{fig:mfclbiomass}.
Selecting a specific biomass time series for an
index of abundance for the MHI stock implicitly assumes
that the ecology and productivity of the index population 
is comparable to the MHI.
On the basis of low levels of exploitation and the location of the
Hawaiian archipelago in the NPTG, MFCL region 2 would seem {\it a
priori} to be the best choice of a for computing an index of
abundance.

\clearpage
\section{Additional Diagnostics}
\label{sec:diagnostics}

\begin{sidewaysfigure}
\begin{center}
{\scriptsize \sffamily
\begin{tabular}{ccc}
\multicolumn{2}{c}{{\small $\sigma_r=0.8$}}\\
Region 2 Index & Region 4 Index\\
\hline
\includegraphics[width=0.30\textwidth]{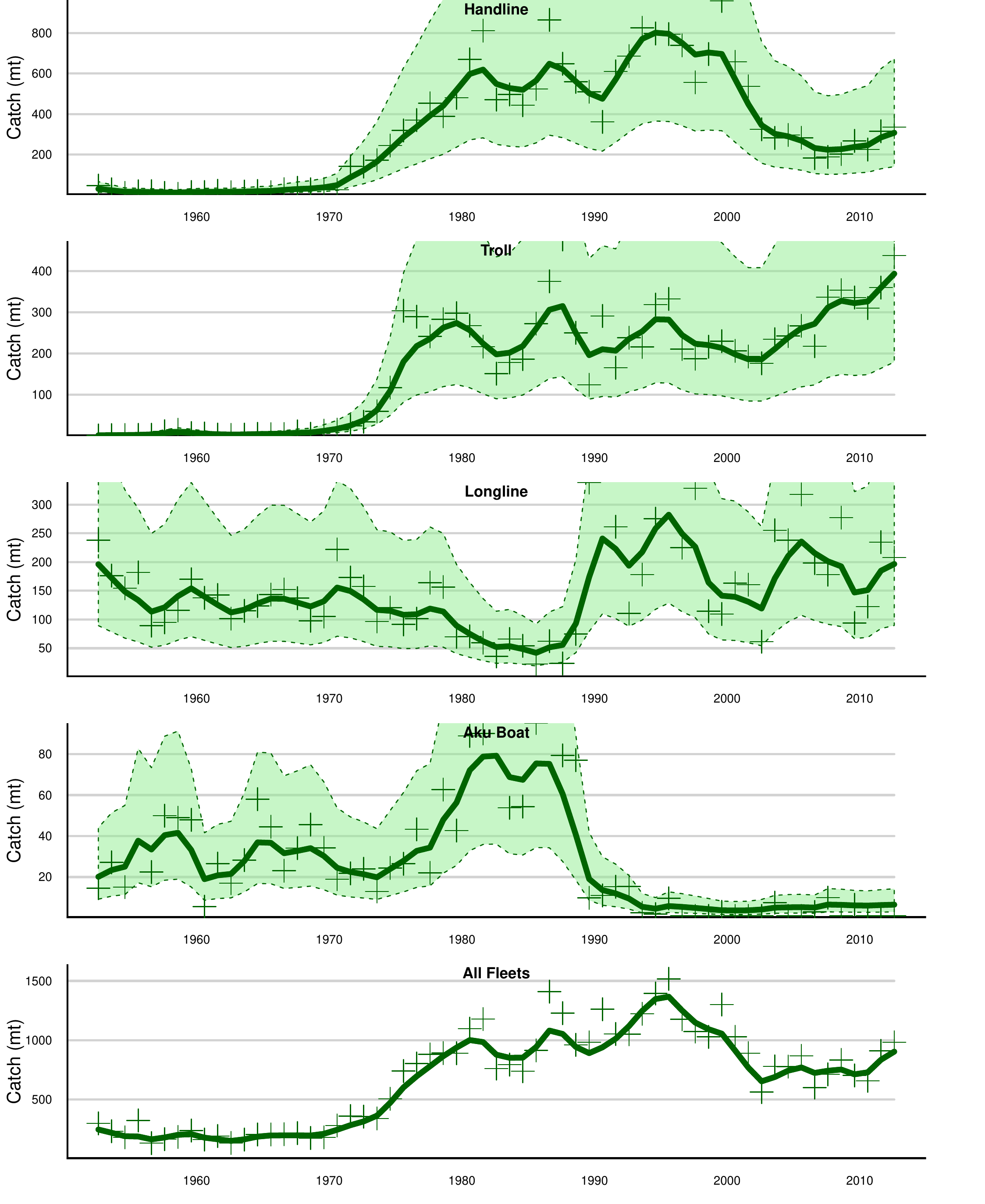} &
\includegraphics[width=0.30\textwidth]{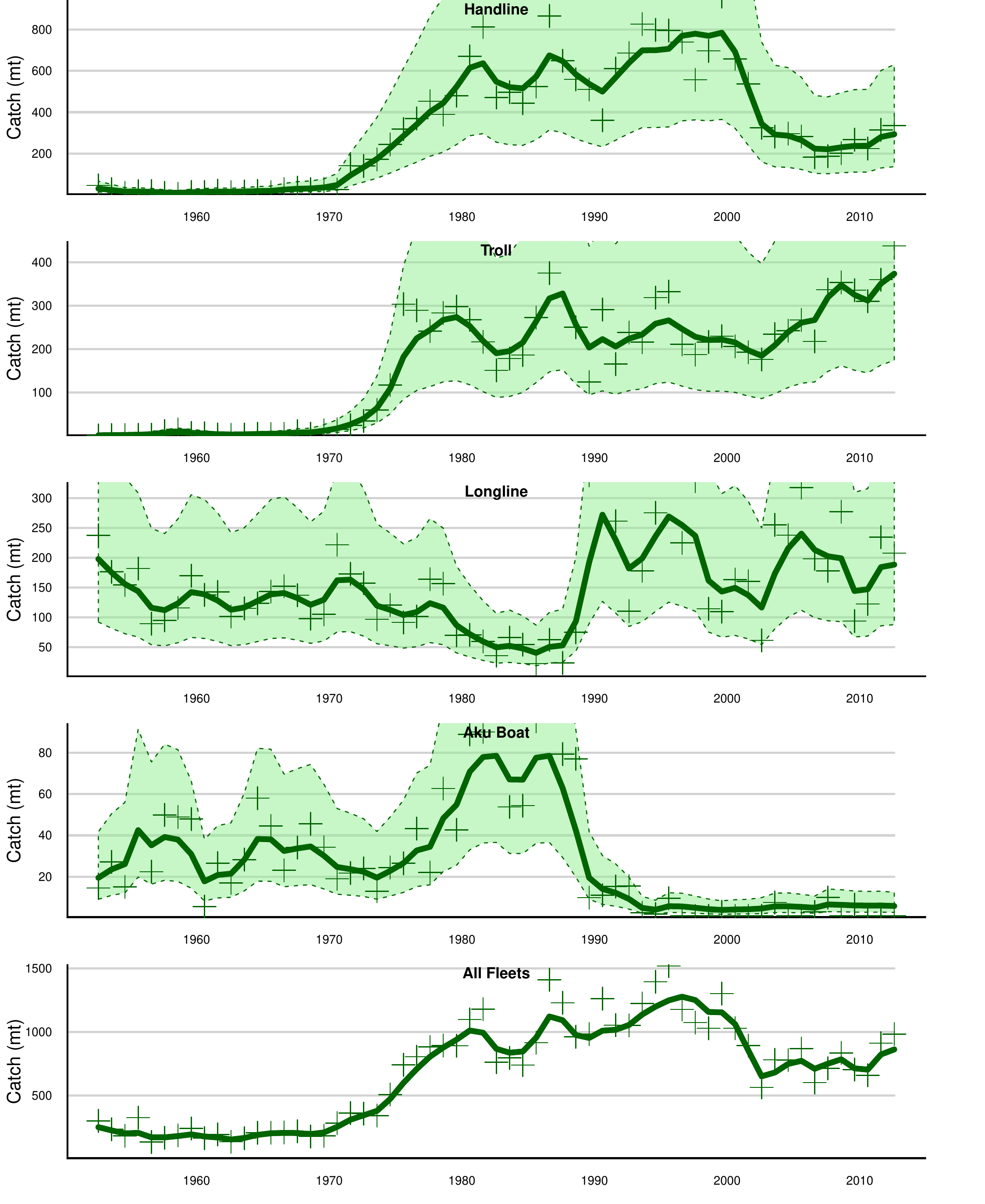} &
\\
\\
\multicolumn{2}{c}{{\small $\sigma_r=0.1$}}\\
Region 2 Index & Region 4 Index\\
\\
\hline
\includegraphics[width=0.30\textwidth]{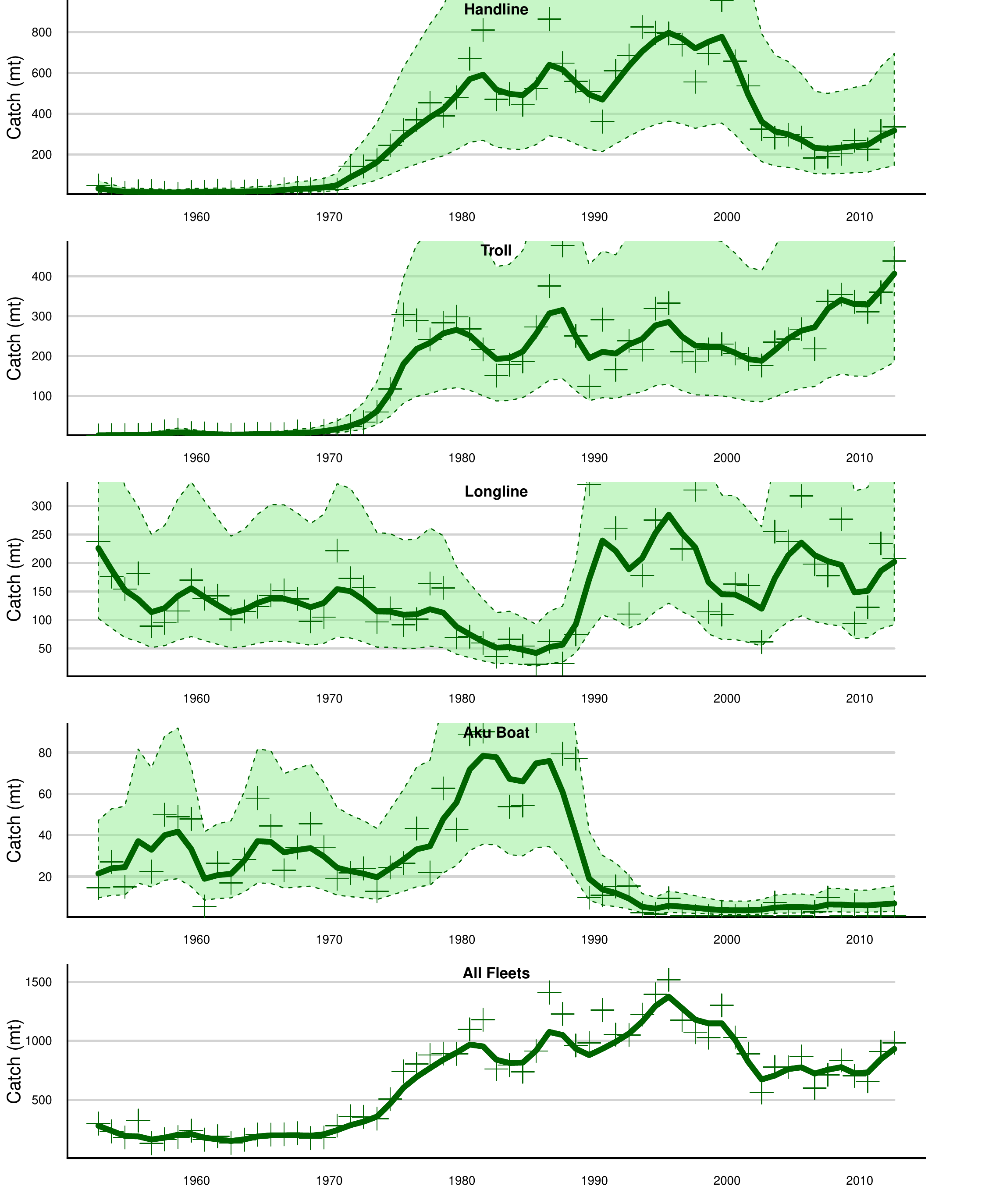} &
\includegraphics[width=0.30\textwidth]{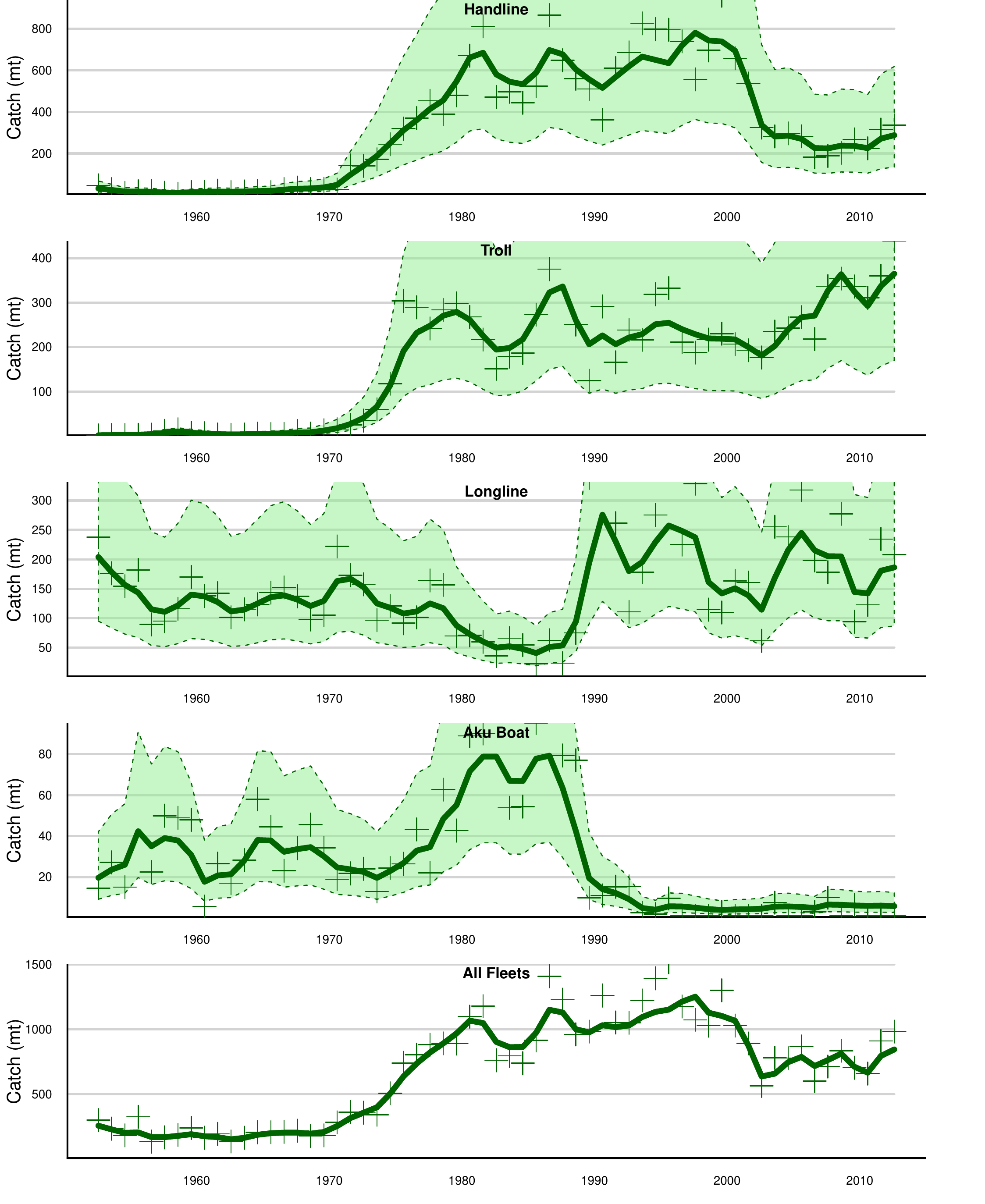} &
\end{tabular}
}
\caption{Trends in observed (+) and estimated catch (solid green
line). The light green shaded areas represent the process error as 
$\pm 2\sigma_Y$.
\label{fig:estC}}
\end{center}
\end{sidewaysfigure}

\begin{sidewaysfigure}
\begin{center}
{\scriptsize \sffamily
\begin{tabular}{ccc}
\multicolumn{2}{c}{{\small $\sigma_r=0.8$}}\\
Region 2 Index & Region 4 Index\\
\hline
\includegraphics[width=0.30\textwidth]{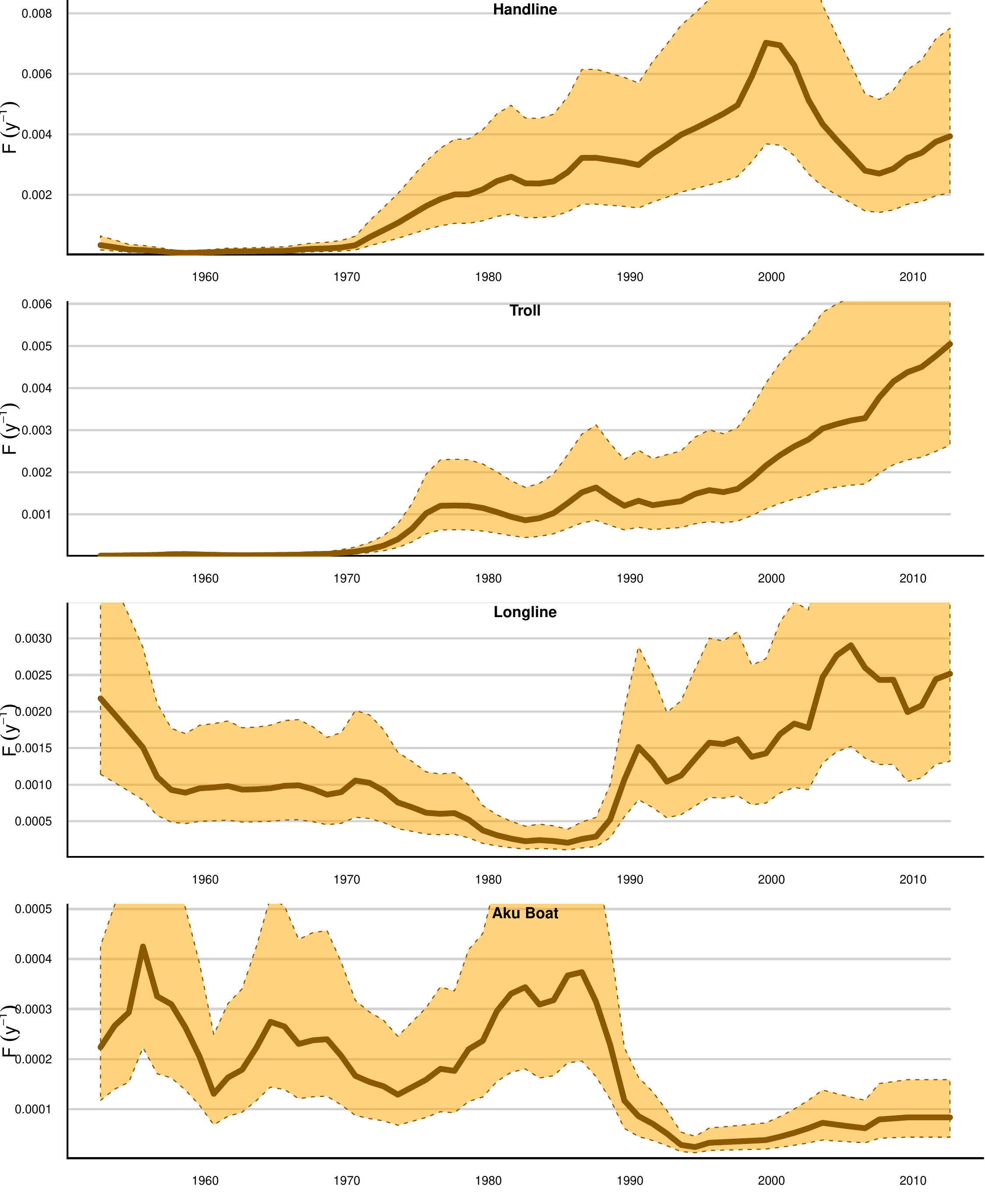} &
\includegraphics[width=0.30\textwidth]{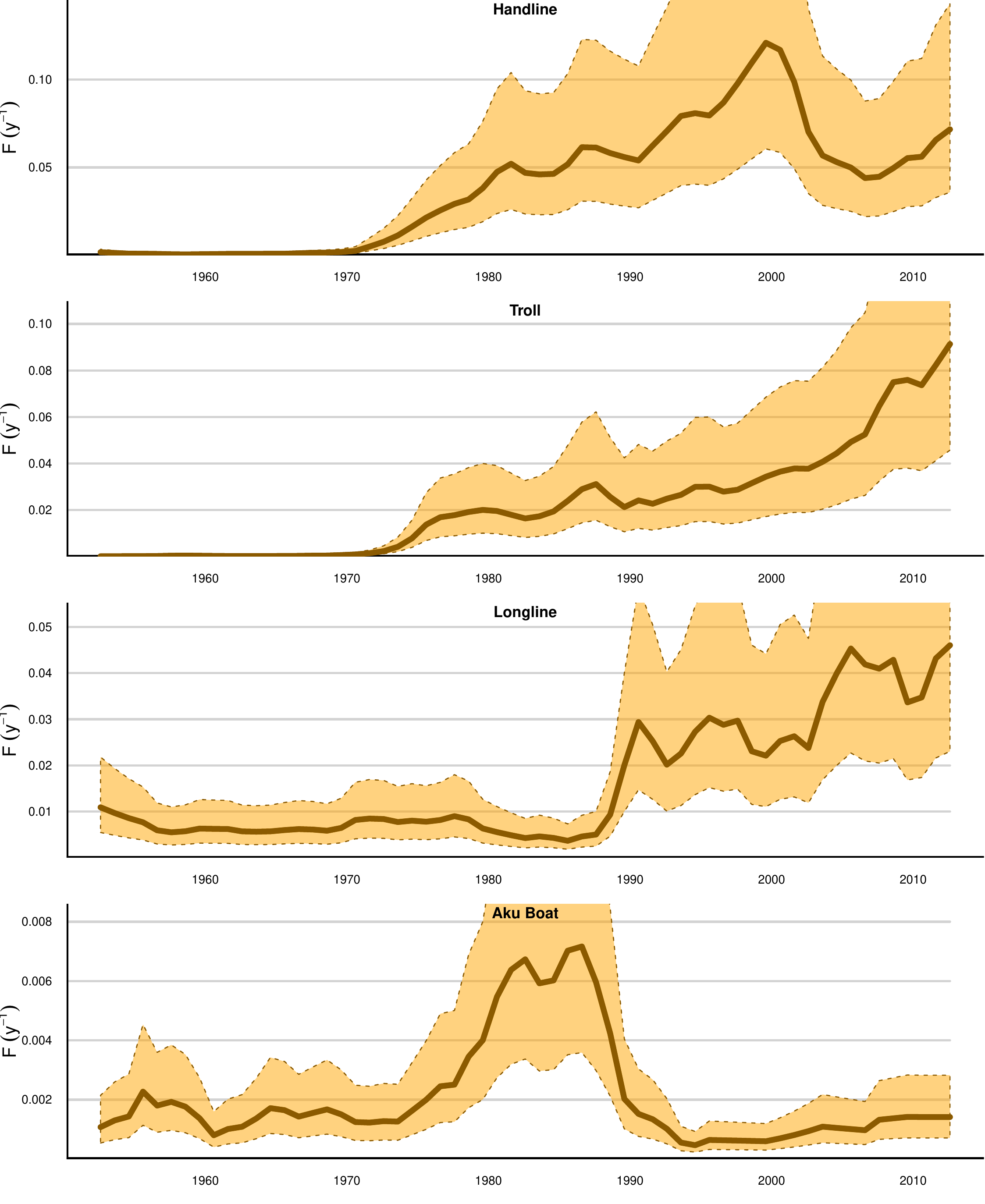} &
\\
\\
\multicolumn{2}{c}{{\small $\sigma_r=0.1$}}\\
Region 2 Index & Region 4 Index\\
\\
\hline
\includegraphics[width=0.30\textwidth]{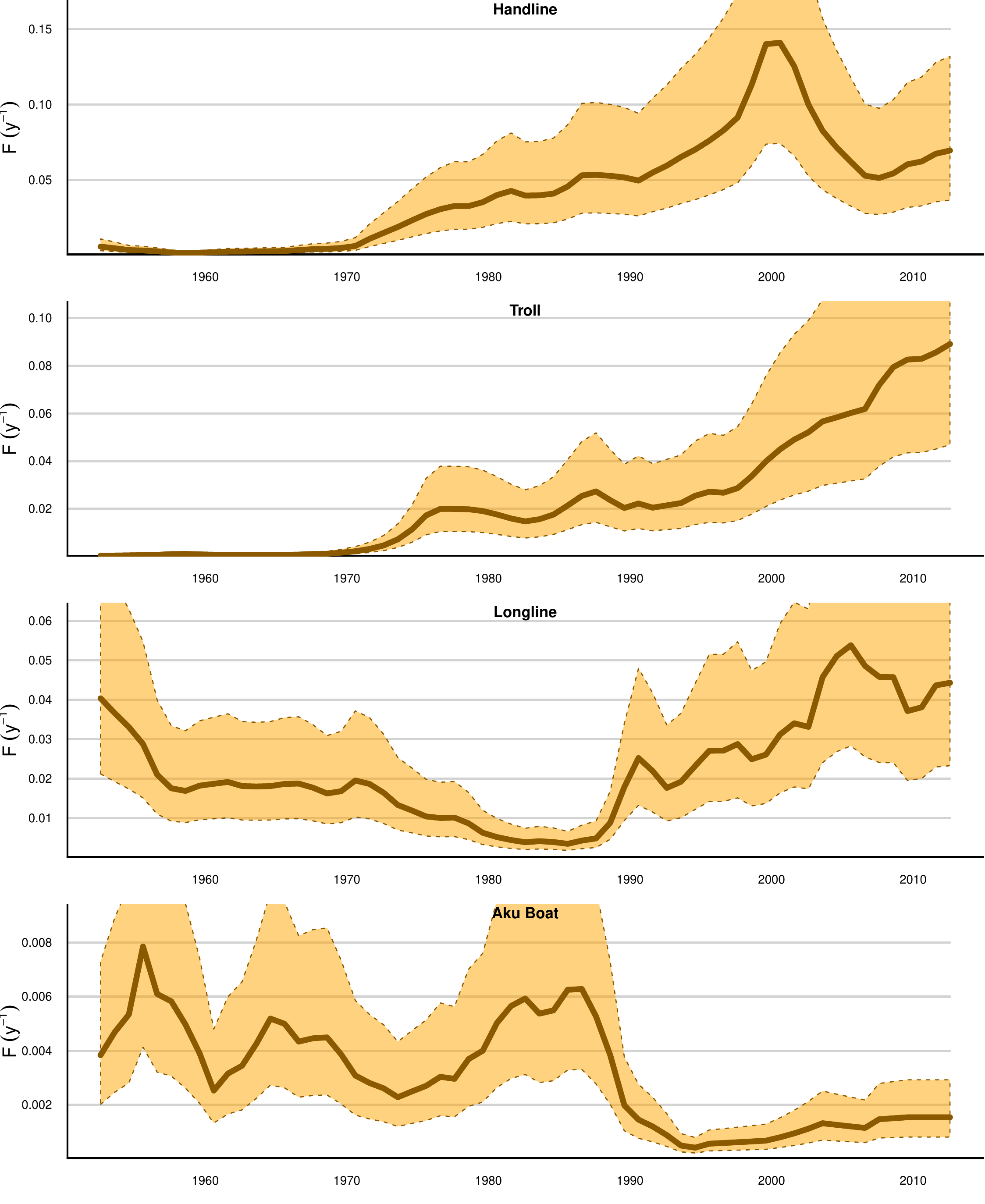} &
\includegraphics[width=0.30\textwidth]{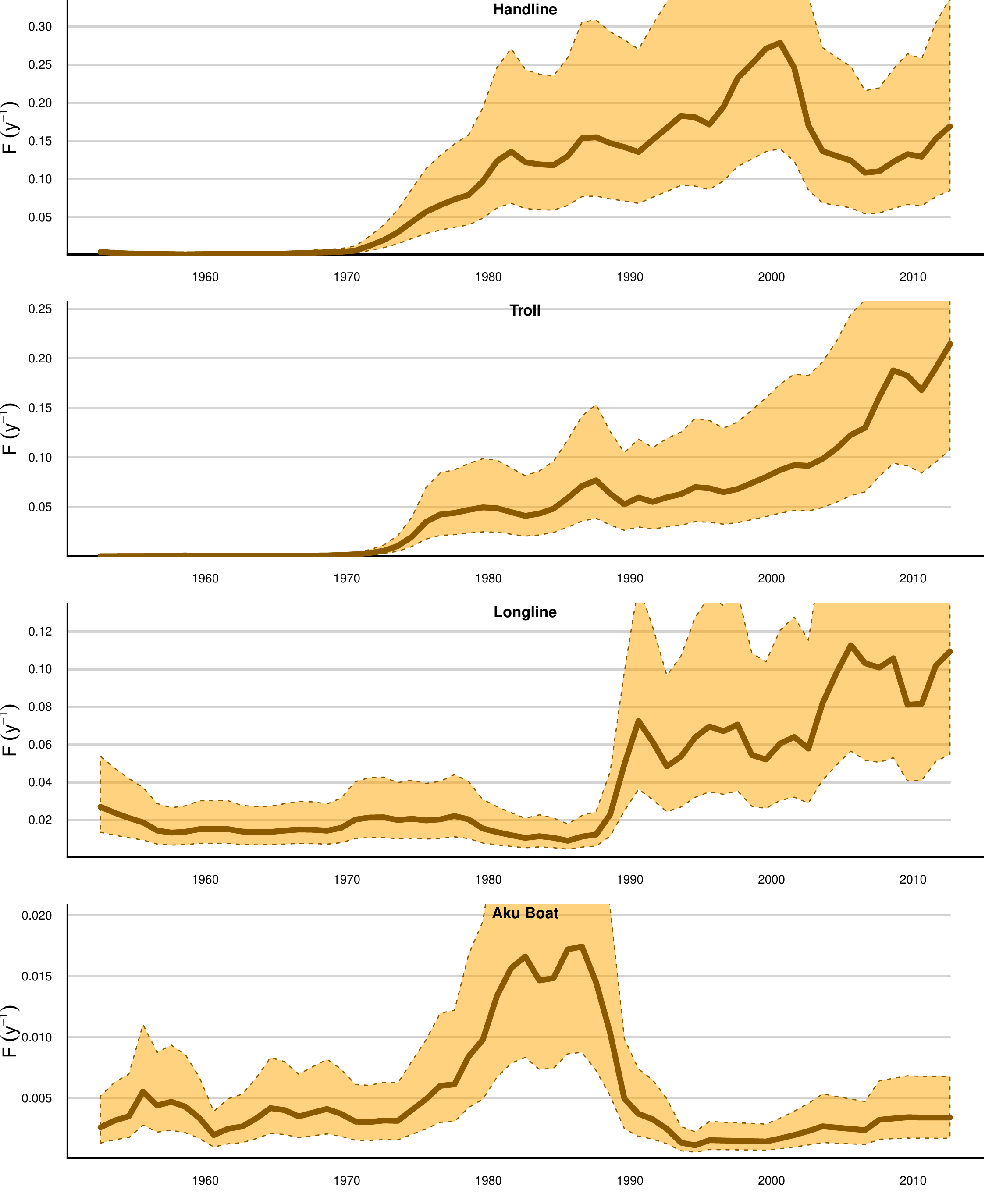} &
\end{tabular}
}
\caption{Trends in estimated fishing mortality (solid brown
line). The orange shaded areas represent the process error as 
$\pm 2\sigma_F$.
\label{fig:estF}}
\end{center}
\end{sidewaysfigure}

Figures~\ref{fig:estC}~and~\ref{fig:estF} show 
the trends in estimated catch add fishing mortality 
with different indices of abundance and values of the $r$
prior, $\sigma_r$.
The ordinate in each panel is scaled to the specific gear type.

\begin{sidewaysfigure}
\begin{center}
{\scriptsize \sffamily
\begin{tabular}{cc}
Region 2; $\sigma_r = 0.8$ & Region 4; $\sigma_r = 0.8$ \\
\includegraphics[width=0.30\textwidth]{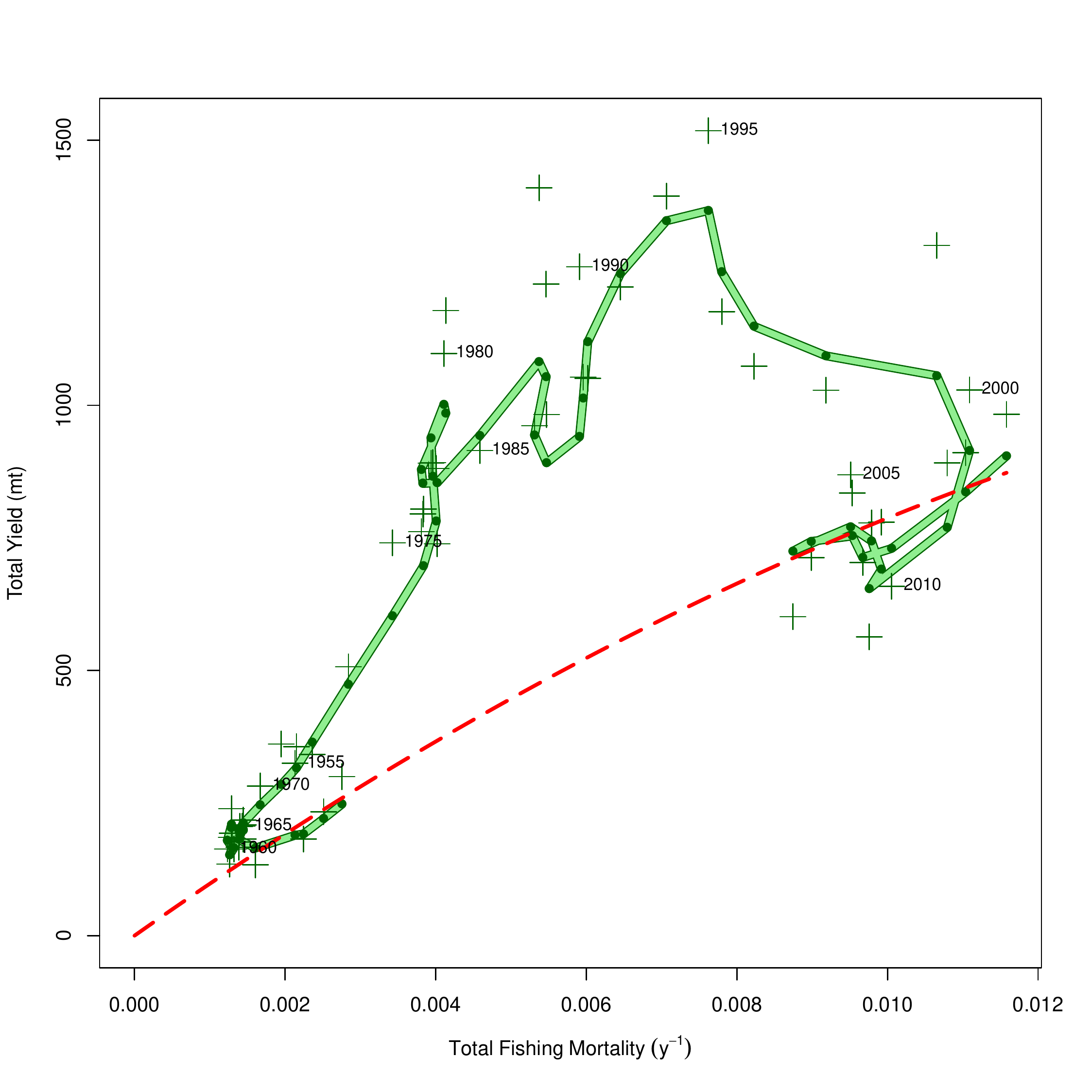} &
\includegraphics[width=0.30\textwidth]{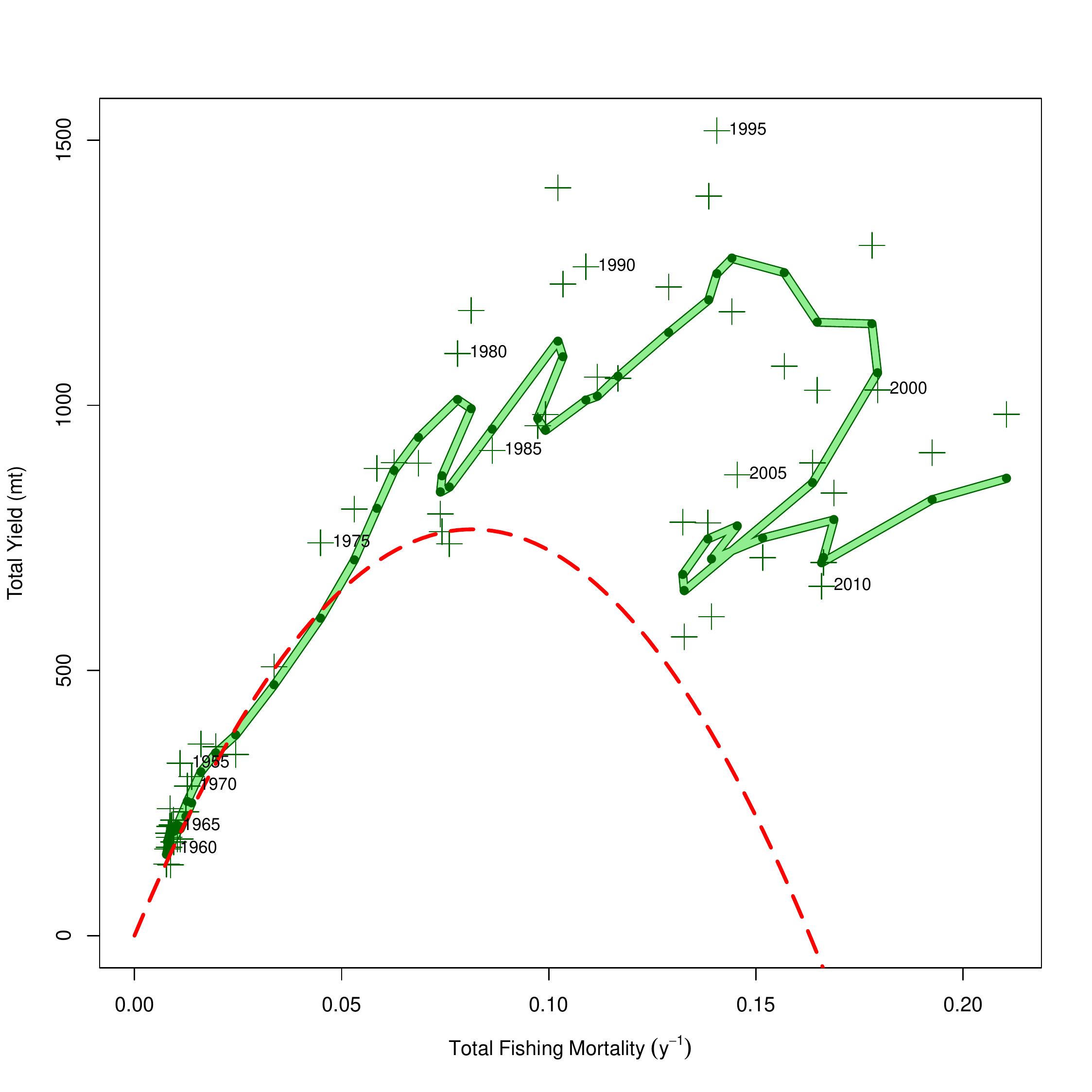} \\
\\
Region 4; $\sigma_r = 0.1$ & Region 4; $\sigma_r = 0.1$ \\
\includegraphics[width=0.30\textwidth]{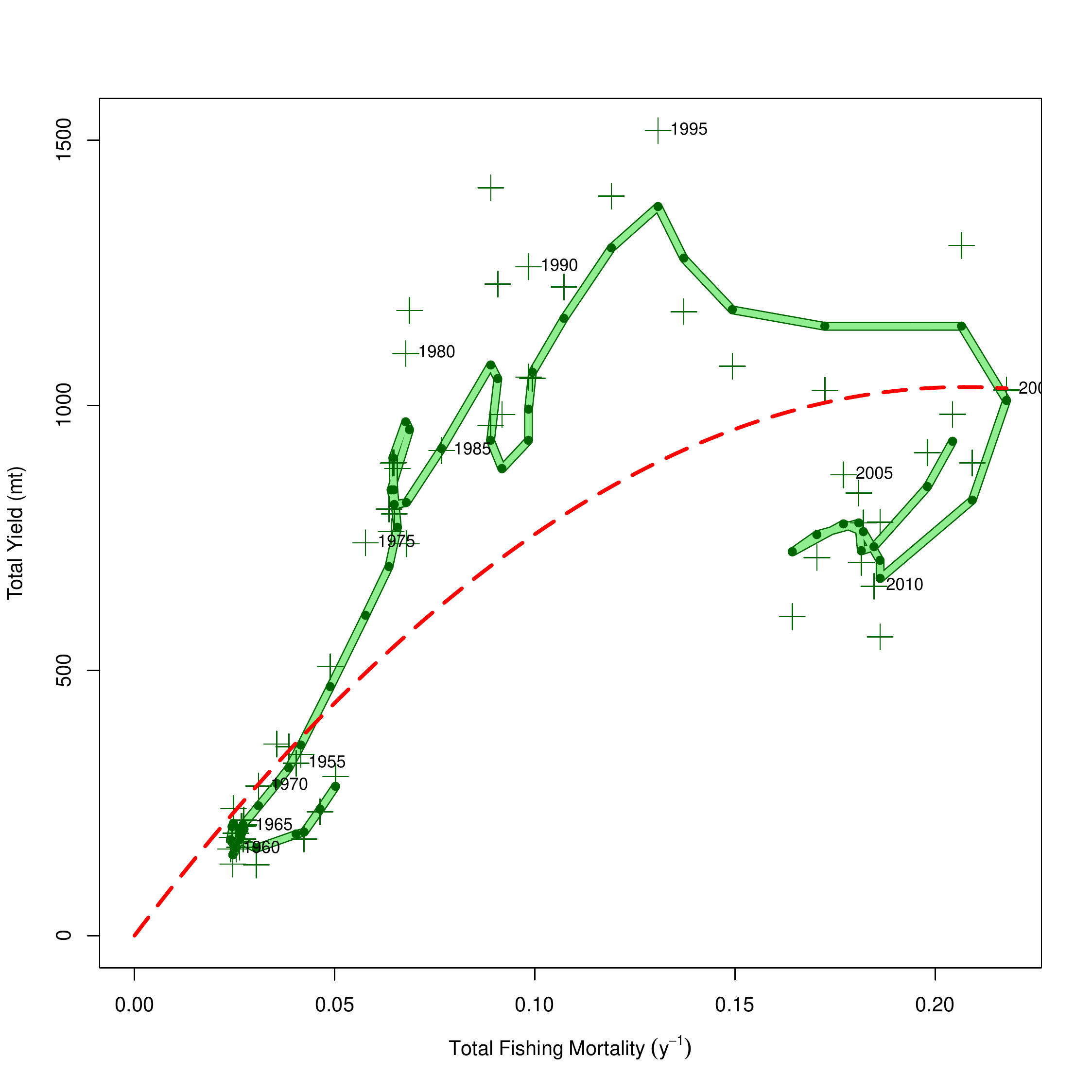} &
\includegraphics[width=0.30\textwidth]{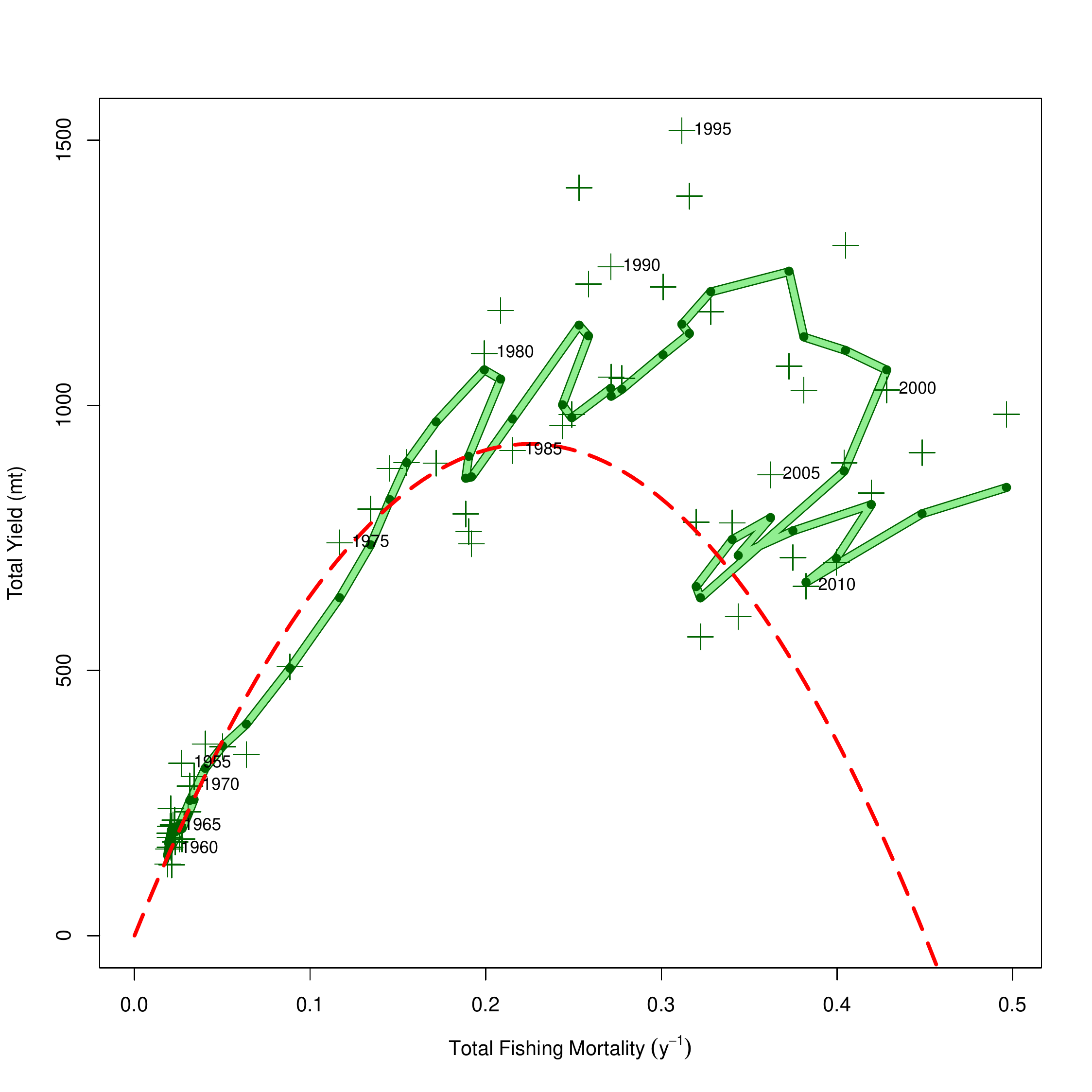} \\
\end{tabular}
}
\caption{Estimated production curves showing catch plotted against
estimated fishing mortality with different indices of abundance
and values of $\sigma_r$.
The green line and dark green dots are estimated catch.
The green $+$ symbols are the observed catch annotated with the year.
The dashed red line is the theoretical equilibrium yield.
Note that the scale of the abscissa is different in each panel, and
the ordinate has been scaled to the observed catch.
\label{fig:estprod}}
\end{center}
\end{sidewaysfigure}

Figure~\ref{fig:estprod} places the model results in a fishery
management context. The parabolic dashed red lines indicate the theoretical
yield from a population with logistic growth. The maximum yield if the
population were at equilibrium would occur at the  peak of the parabola.
There are many possible equilibria, but the  maximum equilibrium yield
has been labeled ``Maximum Sustainable Yield", $\MSY$, in the fisheries
literature.
The parameter $\Fmsy$\ is the fishing mortality that would produce $\MSY$\
at equilibrium. 

\begin{sidewaysfigure}
\begin{center}
{\sffamily
A. Region 2; $\sigma_r=0.8$\\
\includegraphics[width=1.00\textwidth]{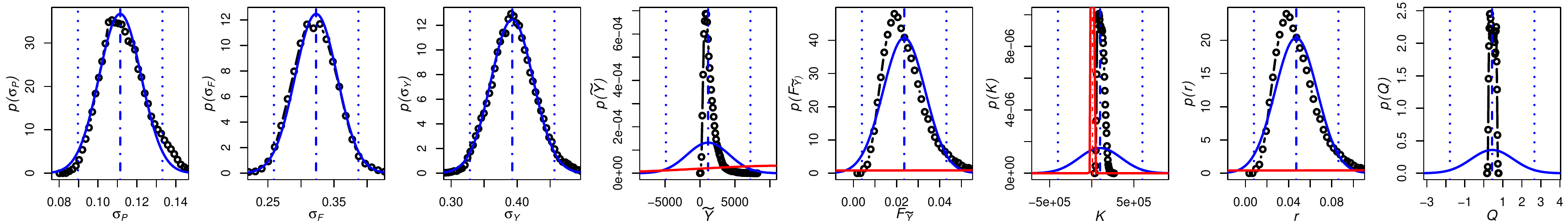}\\
~\\
B. Region 4; $\sigma_r=0.8$\\
\includegraphics[width=1.00\textwidth]{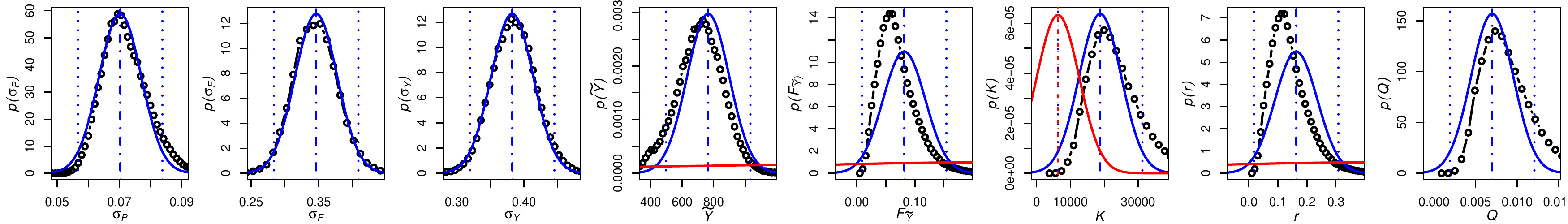}\\
~\\
C. Region 2; $\sigma_r=0.1$\\
\includegraphics[width=1.00\textwidth]{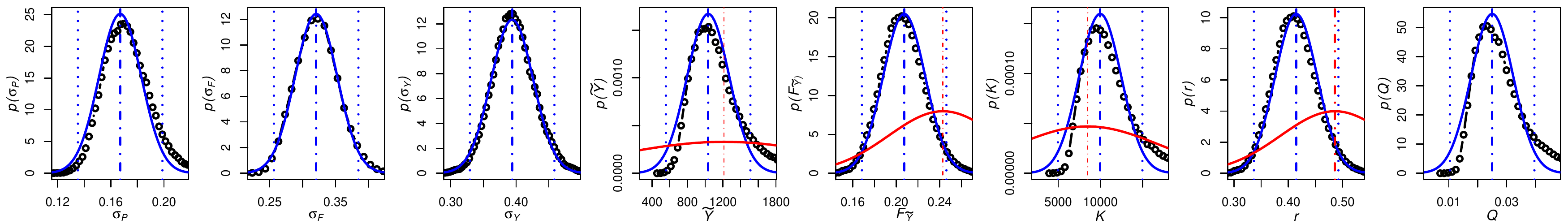}\\
~\\
D. Region 4; $\sigma_r=0.1$\\
\includegraphics[width=1.00\textwidth]{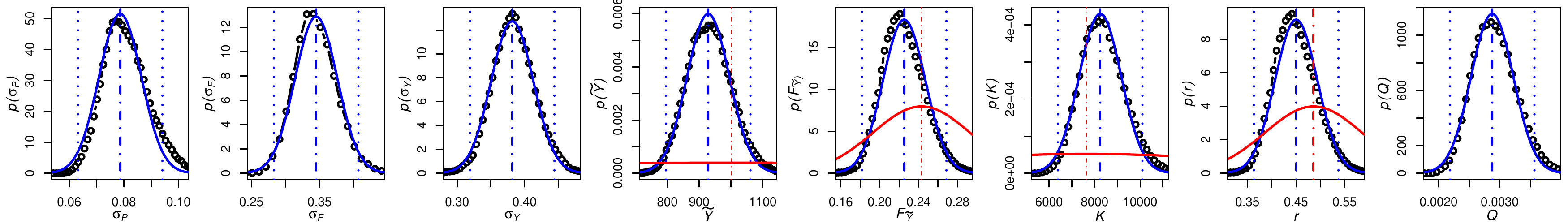}\\
~\\
}
\end{center}
\caption{Posterior distributions of parameter estimates from $10^6$
MCMC samples. 
Black circles and connecting lines are frequencies of the estimates.
Blue curves are the theoretical normal distribution of the maximum
likelihood estimates computed from the standard deviations estimated
from the inverse Hessian matrix at the minimum. 
Vertical blue dotted lines indicate $\pm 2\sigma$.
The solid red curves indicate the effect of the prior on $r$; the
vertical red line is the prior.
\label{fig:posteriors}}
\end{sidewaysfigure}

Figure~\ref{fig:posteriors} shows the frequency distributions of the
parameter estimates obtained by sampling the output of $10^6$ Markov
Chain Monte Carlo simulations. Some parameters are clearly estimated
more accurately than others. The process and observation error
estimates have well defined modes and appear to be normally
distributed around their point estimates. The distributions of other
parameters depends on the abundance index and value of $\sigma_r$.
Figure~\ref{fig:posteriors} also demonstrates the pervasive (some
might say pernicious) effects of using an informative prior on $r$.

\begin{sidewaystable}
\caption{Correlation coefficients between parameter estimates for four
models.
\label{tab:correl}}
   \begin{tabular}[h]{cc}

      \begin{minipage}[t]{0.5\textwidth}
         \scriptsize

\begin{center}
{\normalsize A. Region 2; $\sigma_r = 0.8$}\\
~\\
\begin{tabular}{|lrrrrrrrr|}
\hline
&$\MSY$&$\Fmsy$&$\sigma_P$&$\sigma_F$&$\sigma_Y$&$Q$&$r$&$K$\\
\hline
$\MSY$&1.00&&&&&&&\\
$\Fmsy$&0.23&1.00&&&&&&\\
$\sigma_P$&0.13&0.39&1.00&&&&&\\
$\sigma_F$&0.01&-0.00&-0.00&1.00&&&&\\
$\sigma_Y$&-0.00&0.00&0.00&-0.31&1.00&&&\\
$Q$&0.98&0.04&0.06&0.01&-0.00&1.00&&\\
$r$&0.23&1.00&0.39&-0.00&0.00&0.04&1.00&\\
$K$&0.99&0.07&0.07&0.01&-0.00&1.00&0.07&1.00\\
\hline
\end{tabular}
\end{center}

      \end{minipage} &
      \begin{minipage}[t]{0.5\textwidth}
         \scriptsize

\begin{center}
{\normalsize B. Region 4; $\sigma_r = 0.8$}\\
~\\
\begin{tabular}{|lrrrrrrrr|}
\hline
&$\MSY$&$\Fmsy$&$\sigma_P$&$\sigma_F$&$\sigma_Y$&$Q$&$r$&$K$\\
\hline
$\MSY$&1.00&&&&&&&\\
$\Fmsy$&0.76&1.00&&&&&&\\
$\sigma_P$&0.23&0.28&1.00&&&&&\\
$\sigma_F$&0.04&0.01&-0.02&1.00&&&&\\
$\sigma_Y$&-0.02&-0.01&-0.00&-0.26&1.00&&&\\
$Q$&-0.53&-0.95&-0.25&0.01&0.01&1.00&&\\
$r$&0.76&1.00&0.28&0.01&-0.01&-0.95&1.00&\\
$K$&-0.50&-0.94&-0.26&0.01&0.01&0.99&-0.94&1.00\\
\hline
\end{tabular}
\end{center}

      \end{minipage}\\
      & \\
      & \\
      \begin{minipage}[t]{0.5\textwidth}
         \scriptsize

\begin{center}
{\normalsize C. Region 2; $\sigma_r = 0.1$}\\
~\\
\begin{tabular}{|lrrrrrrrr|}
\hline
&$\MSY$&$\Fmsy$&$\sigma_P$&$\sigma_F$&$\sigma_Y$&$Q$&$r$&$K$\\
\hline
$\MSY$&1.00&&&&&&&\\
$\Fmsy$&-0.01&1.00&&&&&&\\
$\sigma_P$&0.08&0.33&1.00&&&&&\\
$\sigma_F$&0.05&-0.01&-0.02&1.00&&&&\\
$\sigma_Y$&-0.02&0.01&-0.01&-0.30&1.00&&&\\
$Q$&0.90&-0.38&-0.03&0.05&-0.02&1.00&&\\
$r$&-0.01&1.00&0.33&-0.01&0.01&-0.38&1.00&\\
$K$&0.93&-0.39&-0.05&0.05&-0.02&0.98&-0.39&1.00\\
\hline
\end{tabular}
\end{center}

      \end{minipage} &
      \begin{minipage}[t]{0.5\textwidth}
         \scriptsize

\begin{center}
{\normalsize D. Region 4; $\sigma_r = 0.1$}\\
~\\
\begin{tabular}{|lrrrrrrrr|}
\hline
&$\MSY$&$\Fmsy$&$\sigma_P$&$\sigma_F$&$\sigma_Y$&$Q$&$r$&$K$\\
\hline
$\MSY$&1.00&&&&&&&\\
$\Fmsy$&0.13&1.00&&&&&&\\
$\sigma_P$&0.07&0.17&1.00&&&&&\\
$\sigma_F$&0.08&-0.01&-0.04&1.00&&&&\\
$\sigma_Y$&-0.04&0.00&-0.01&-0.26&1.00&&&\\
$Q$&0.49&-0.76&-0.08&0.06&-0.03&1.00&&\\
$r$&0.13&1.00&0.17&-0.01&0.00&-0.76&1.00&\\
$K$&0.53&-0.78&-0.10&0.06&-0.03&0.96&-0.78&1.00\\
\hline
\end{tabular}
\end{center}
      \end{minipage}\\
   \end{tabular}
\end{sidewaystable}

\clearpage
\section{Noncommercial Catch}
\label{sec:klingon}
The Main Hawaiian Islands support a large number of noncommercial
fishers. There is no catch reporting system and no mandatory marine fishing
license. 
Accurately estimating the noncommercial catch is therefore difficult. 
The State of Hawaii Department of Land and Natural Resources and the
United States National Oceanic and Atmospheric Administration are  
collaborating to improve survey methods for the Hawai`i Marine
Recreational Fishing Survey (HMRFS). The HMRFS Newsletter (DLNR, 2011)
reports that non-commercial catches of YFT range between 2300 and 6900
mt from 2003 through 2010,  roughly 3 to 10 times greater than the most recent
commercial catch. The assessment model developed here has the capability to
augment the reported catch with catch by an additional fleet in
proportion to the total reported catch and use the augmented catch in
the assessment as if it were reported. Region 2 indexing was used with
both loose and tight priors on $r$.  Four different multipliers
were tested so that the augmented catch was 1, 2, 5, and 10 times total catch.

Tables~\ref{tab:k2} and \ref{tab:k2-rprior} summarize the results of
fitting the augmented catch.
All 8 augmented catch models converge to solutions. The negative
log likelihood values are higher than comparable non-augmented fits.
{\it A priori} one would expect that a fish population capable
of producing a higher than reported yield over a 60 year history
would have a higher productivity and a higher biomass. These results
are generally consistent with this conclusion. 
The model compensates for the higher catch by estimating higher
$\MSY$ and higher $K$ in all models, 
whereas estimates $\Fmsy$\ and $r$ are relatively unchanged by the
augmented catch.
Estimates of $\bar{F}_5$ increase in models with $\sigma_r=0.8$ but
change very little in models with $\sigma_r=0.1$.
The $\sigma_r=0.8$ model estimates that the fishery is sustainable
with a total yield about 5 times the current reported yield.
The $\sigma_r=0.1$ model forces a much higher $\Fmsy$. In consequence,
this form of the model estimates a sustainable fishery with yields at
least 10 times current reported yield. 

This simple catch augmentation simulation indicates that unreported
noncommercial catch causes this model to underestimate maximum
sustainable yield and the fishing mortality at maximum sustainable
yield when applied to the available reported catch data. To put this
conclusion in a different context, fishers should report their catch
if they want a higher yield from a managed fishery.

\begin{table}
{\small
\caption{Results of fitting the model to 4 different multiples of the
total catch using MFCL Region 2 as an index of abundance and a loose
prior on $r, \sigma_r = 0.8$, comparable to Table~\ref{tab:ests1}. 
See Table~\ref{tab:allvars1} for definitions of all variables.
\label{tab:k2}}
\begin{center}
\begin{tabular}{|l|rrrr|}
\hline
Multiplier & 1 & 2 & 5 & 10\\
\hline
&k1 & k2 & k5 & k10\\
$-\log L$ & 181.75 & 181.77 & 181.97 & 182.84\\
$n$ & 6 & 6 & 6 & 6\\
$|G|_{max}$ & 8.32e-05 & 6.87e-05 & 6.24e-05 & 5.6e-05\\
\hline
$\MSY$ & 2240 & 2740 & 2810 & 3070\\
$\Fmsy$ & 0.0237 & 0.0237 & 0.0237 & 0.0248\\
$Q$ & 0.819 & 1 & 1 & 1\\
$\sigma_P$ & 0.112 & 0.112 & 0.112 & 0.113\\
$\sigma_F$ & 0.295 & 0.295 & 0.295 & 0.294\\
$\sigma_Y$ & 0.358 & 0.358 & 0.358 & 0.358\\
\hline
$r$ & 0.0474 & 0.0474 & 0.0475 & 0.0495\\
$K$ & 189000 & 232000 & 237000 & 248000\\
$\bar{F}_5$ & 0.0111 & 0.0136 & 0.0270 & 0.0485\\
\hline
$p_0$ & 0.197 & 0.197 & 0.197 & 0.197\\
$\tilde{r}$ & 0.486 & 0.486 & 0.486 & 0.486\\
$\sigma_r$ & 0.8 & 0.8 & 0.8 & 0.8\\
$\bar{Y}_5$ & 1630 & 2450 & 4890 & 8960\\
\hline
\end{tabular}
\end{center}
}
\end{table}

\begin{table}
{\small
\caption{Results of fitting the model to 4 different multiples of the
total catch using MFCL Region 2 as an index of abundance and a tighter
prior on $r, \sigma_r = 0.1$, comparable to Table~\ref{tab:ests1}. 
See Table~\ref{tab:allvars1} for definitions of all variables.
\label{tab:k2-rprior}}
\begin{center}
\begin{tabular}{|l|rrrr|}
\hline
Multiplier  & 1 & 2 & 5 & 10\\
\hline
&k1 & k2 & k5 & k10\\
$-\log L$ & 194.64 & 194.63 & 194.58 & 194.56\\
$n$ & 6 & 6 & 6 & 6\\
$|G|_{max}$ & 9.36e-05 & 4.46e-05 & 2.7e-05 & 7.46e-05\\
\hline
$\MSY$ & 2060 & 3110 & 6260 & 11500\\
$\Fmsy$ & 0.208 & 0.208 & 0.208 & 0.208\\
$Q$ & 0.0495 & 0.0747 & 0.15 & 0.277\\
$\sigma_P$ & 0.169 & 0.169 & 0.167 & 0.167\\
$\sigma_F$ & 0.291 & 0.291 & 0.291 & 0.291\\
$\sigma_Y$ & 0.357 & 0.357 & 0.357 & 0.358\\
\hline
$r$ & 0.415 & 0.415 & 0.416 & 0.416\\
$K$ & 19800 & 29900 & 60200 & 111000\\
$\bar{F}_5$ & 0.191 & 0.191 & 0.191 & 0.191\\
\hline
$p_0$ & 0.197 & 0.197 & 0.197 & 0.197\\
$\tilde{r}$ & 0.486 & 0.486 & 0.486 & 0.486\\
$\sigma_r$ & 0.1 & 0.1 & 0.1 & 0.1\\
$\bar{Y}_5$ & 1630 & 2450 & 4890 & 8960\\
$\bar{C}$ & 1380 & 2070 & 4130 & 7560\\
\hline
\end{tabular}
\end{center}
}
\end{table}

\end{document}